%% file: vienna-arxiv.tex
\documentclass[namedreferences]{solarphysics}
\usepackage[hyperref,optionalrh,solaromanenum]{spr-sola-addons_arxiv} 
\usepackage{graphicx} 
\usepackage{color}
\usepackage{breakurl}
\usepackage{acronym}

\hypersetup{
  colorlinks=true,   
  urlcolor=blue,     
  linkcolor=blue,    
}

\usepackage{twoopt}  

\def\figs{.}

\input{rrmacros-pub}

\def\revapar{}                 
\long\def\reva#1{#1}           
\def\revbpar{}                 
\long\def\revb#1{#1}           
\def\revcpar{}                 
\long\def\revc#1{#1}           
\long\def\revd#1{#1}           

\def\linkrtsa#1#2{\linkadspage{2003rtsa.book.....R}{#1}{RTSA #2}}
\def\acp#1{#1} 

\def\rmit#1{{\it #1}}         
\def\eg{\rmit{e.g.}}  
\bibpunct{(}{)}{;}{a}{,}{,}   
\def\specchar#1{{\sc{#1}}}    

\begin{document}

\begin{article}

\begin{opening}

\title{Non-Equilibrium Spectrum Formation Affecting Solar Irradiance}

\author[addressref={aff1,aff2,aff3},corref,email={R.J.Rutten@uu.nl}]
{\inits{R.J.~}\fnm{Robert~J.~ }\lnm{Rutten}} 

\address[id=aff1]{\LA}
\address[id=aff2]{\ITA}
\address[id=aff3]{\RoCS}

\runningauthor{Robert J. Rutten}
\runningtitle{Non-Equilibrium Affecting Irradiance}

\begin{abstract}
This is an \revapar overview of non-equilibrium aspects of the
formation of solar continua and lines affecting the contributions by
\reva{magnetic} network and plage to spectrally resolved solar
irradiance.
After a \reva{brief summary} of these contributions and a compact
\reva{refresher} of solar spectrum formation the emphasis is on
graphical exposition.
\reva{Major} obstacles for \revapar simulation-based irradiance studies are
how to cope with NLTE scattering in the violet and ultraviolet line
haze and how to cope with \reva{retarded} hydrogen opacities in
infrared and mm radiation. 
\revapar
\end{abstract}
\keywords{Spectrum, Solar Irradiance}
\end{opening}

\section{Introduction} \label{sec:introduction} 

A major contributor to solar irradiance variability are the small but
ubiquitous kilogauss magnetic concentrations (henceforth MC) that
constitute solar network and plage.
They are obvious in any longitudinal magnetogram from the {\em
Helioseismic and Magnetic Imager} of the {\em Solar Dynamics
Observatory} as bipolar salt-and-pepper grains spread in roughly
cellular patterns \reva{(network) and denser unipolar patches \reva{of
grains} in or near active regions (plage)}.
Towards the limb their facular presence is clearer as bright grains in
1700\,\AA\ images from the {\em Atmospheric Imaging Assembly} of the
{\em Solar Dynamics Observatory\/}.\footnote{\revc{Nomenclature: MC
patches were traditionally recognized on \CaIIHK\ spectroheliograms
and called flocculi (Hale and Ellerman) and plage (Deslandres). 
Their chromospheric appearance is coarser, hence more evident, 
than \revd{underlying} photospheric MCs but corresponds closely to the
MC surface patterning as network and plage.
The 1700\,\AA\ grains are photospheric.}}

Limiting this overview to \acp{MC} spectrum formation ignores the
irradiance variability contributions of sunspots, filaments, and
flares. 
In contrast to these, \acp{MC}s are nowadays well reproduced in
numerical \acp{MHD} simulations.
\reva{Presently,} solar irradiance modeling of their contribution
\reva{progresses} from static 1D temperature-stratification fitting to
3D$(t)$ simulation-based interpretation (\eg\
\citeads{2017A&A...605A..45N}). 
Handling non-equilibrium spectrum formation in this transition is
mandatory but nontrivial. 

The primary radiation mechanism by which \acp{MC}s \reva{are brighter
than their surroundings is not temperature enhancement as proposed
originally by} \citetads{1970SoPh...14..315C} 
and \citetads{1975SoPh...42...79S}. 
\reva{Instead, it} is enhanced ``hole-in-the-surface'' radiation due to
the Wilson depressions \reva{in} MCs, \reva{amounting to} a few hundred
km, \reva{which result} from partial evacuation by the magnetic
pressure of their kilogauss fields \reva{found} by
\citetads{1978A&A....70..789F} 
with Stenflo's
(\citeyearads{1973SoPh...32...41S}) 
line-ratio technique.
The magnetostatic thin-fluxtube model of
\citetads{1976SoPh...50..269S} 
inspired by
\citetads{1967SoPh....1..478Z} 
\reva{explained} such hole radiation. 
\reva{It} was substantiated in detail by Solanki and coworkers (review
by \citeads{1993SSRv...63....1S}) 
and verified with time-dependent \acp{MHD} simulations in 2D (\eg\
Grossmann-Doerth \etal\
\citeyearads{1994A&A...285..648G}, 
\citeyearads{1998A&A...337..928G}; 
\citeads{1998ApJ...495..468S}; 
\citeads{2001SoPh..203....1G}) 
and then in 3D
(\citeads{2004ApJ...607L..59K}; 
\citeads{2004ApJ...610L.137C}; 
\citeads{2005A&A...429..335V}; 
\citeads{2009A&A...504..595Y}). 

Many spectral features gain extra brightness from extra \acp{MC}
evacuation: filigree and line gaps seen in minority-stage lines (\eg\
\FeI) from extra ionization, bright \reva{grains} in ultraviolet
continua likewise from extra minority-stage ionization, in the CN band
around 3883\,\AA\ and the CH G-band around 4305\,\AA\ from extra
dissociation, in the outer wings of the Balmer lines, \revc{of} \CaII\
\HK, and \revc{of} \CaIR\ from less collisional damping.
Of these \revd{the blue} \Halpha\ \revd{wing} is the brightening
champion
(\citeads{2006A&A...452L..15L}). 
\MnI\ lines are special in lacking Doppler sensitivity to the
surrounding granulation\reva{,} darkening that instead
(\citeads{2009A&A...499..301V}). 

\reva{Because MCs are small features these brightenings became known as
``bright points''.
MC} bright-point observation needs sufficient angular resolution to
avoid \revapar cancelation by smearing with the darkness of the
intergranular lanes \revapar 
in which \acp{MC}s reside
(\citeads{1996ApJ...463..797T}). 
With the superior resolution of the Swedish 1-m Solar Telescope (SST)
G-band bright points were resolved into more intricate morphologies
(\citeads{2004A&A...428..613B}), 
but \revb{these also} \revapar \revb{brighten by} hole-in-the-surface
\revb{radiation}. \revapar \reva{Limbward faculae gain stalk-like
brightness} from deeper penetration into hot granules behind the
\acp{MC}s through \reva{the MC} opacity gap along the line of sight
and \reva{also gain} dark \reva{feet} from the MC-surrounding lanes
(\linkadspage{1999ASPC..184..181R}{8}{Figure~7} of
\citeads{1999ASPC..184..181R}; 
\citeads{2005A&A...430..691S}). 

\revapar Plage and network irradiance modeling has generally ignored
\reva{this} \revapar multi-D nature of MC hole brightening by
reverting to \reva{classic} 1D description, particularly in the {\em
Spectral And Total Irradiance REconstruction} (\textsf{SATIRE})
\reva{efforts} initiated by
\citetads{1999A&A...345..635U}. 
\reva{\revc{These} employ} the plage model of
\citetads{1993ApJ...406..319F} 
after undoing its chromosphere to permit \acp{LTE} line synthesis with
the \textsf{ATLAS} code of Kurucz
(\citeyearads{1970SAOSR.309.....K}, 
\citeyearads{1993KurCD..13.....K}) without getting strong-line core
reversals.
The same tactic was used in hundreds of \acp{LTE} abundance
\reva{studies} relying on the chromosphere-less 1D model of
\citetads{1974SoPh...39...19H} 
until the 3D$(t)$ \acp{NLTE} revolution \reva{in abundance
determination}
(\citeads{2009ARA&A..47..481A}). 
Such 1D plage models necessarily have increasing temperature excess
over their quiet-Sun companions because in 1D modeling actual
hole-in-the surface brightening requires this simulacrum, with the
divergence starting already in the deep photosphere although there is
no observational evidence of \acp{MC} heating in the first
\reva{hundreds of kilometers} in standard height
(\citeads{2005A&A...437.1069S}). 

\revapar Twenty years ago the \textsf{SATIRE} 1D and \acp{LTE}
assumptions were forgivable trac\-ta\-bility ones \revapar
\revc{permitting} \acp{MC} irradiance modeling throughout the
spectrum, but with present computer resources they need to be relaxed
or at least verified similarly to the \revapar abundance
revolution. \revapar
\citetads{2017A&A...605A..45N} 
went from static 1D to time-dependent 3D but kept LTE synthesis with
\textsf{ATLAS}.
Adding non-equilibrium spectral synthesis is next. 
Fortunately solar spectrum formation in and around \acp{MC}s does not
differ intrinsically from spectrum formation in non-magnetic areas or
even in idealized 1D static atmospheres.  
The basic \revc{radiation} physics is the same; all lessons learned in
past decades apply.
The major extension is from static 1D to time-dependent 3D geometry
and \reva{corresponding} numerical complexities and challenges.  

Sections~\ref{sec:equilibria}\,--\,\revb{\ref{sec:schemscat}}
\reva{below} summarize \revapar basic lessons, \reva{with frequent
reference\footnote{\reva{Including page links: depending on your pdf
viewer and its settings, the links to specific pages of RTSA and other
ADS-available publications may open the pertinent page directly in
your browser.
Clicking on citations may open the corresponding \acp{ADS} abstract
page (not in the Springer-mutilated published version).}}
to my masters-level lecture notes
(\citeads{2003rtsa.book.....R}, 
henceforth RTSA) to avoid repeating treatments given there, while
expanding on these by adding more recent teaching
material}\footnote{Since the
\href{http://www.staff.science.uu.nl/~rutte101/Closure_Utrecht.html}
{closure of Utrecht astronomy} I welcome invitations to teach solar
spectrum formation at masters level
\href{http://www.staff.science.uu.nl/~rutte101/My_courses.html}{elsewhere}
-- taking a week for what is summarized here.}
from \href{http://www.staff.science.uu.nl/~rutte101} {my
website}\footnote{\reva{\url{www.staff.science.uu.nl/~rutte101}. 
If defunct search ``Rob Rutten webstek''.}}, including lecture
displays from {\em Solar Spectrum Formation: Theory}
(\href{http://www.staff.science.uu.nl/~rutte101/rrweb/rjr-edu/lectures/rutten_ssf_lec.pdf}{SSF})
and {\em Solar Spectrum Formation: Examples}
(\href{http://www.staff.science.uu.nl/~rutte101/rrweb/rjr-edu/lectures/rutten_ssx_lec.pdf}{SSX})
\reva{found there under
\href{http://www.staff.science.uu.nl/~rutte101/Astronomy_course.html}{``Astronomy
course material''}.}

\revb{Section~\ref{sec:demonstrations} uses NLTE synthesis of selected
continua and lines in a 1D atmosphere to illustrate NLTE effects
affecting irradiance.}

Section~\ref{sec:obstacles} discusses major \revapar 
\reva{non-equilibrium} obstacles in numerical modeling of the
irradiance contributions by network and plage.

\bibnote{2003rtsa.book.....R}{(RTSA)}

\begin{figure}
  \centering
  \includegraphics[width=12cm]{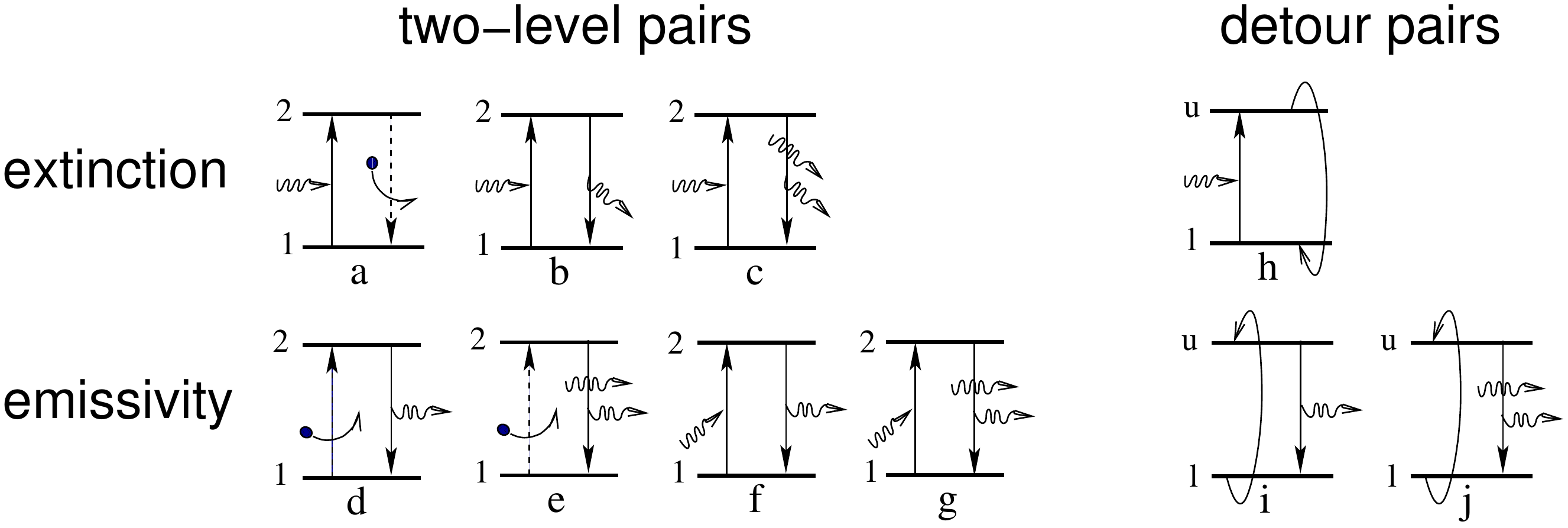}
  \caption[]{
  Atomic transitions governing line formation arranged in
  photon-involving pairs. \revapar The beam of interest (direction of
  the intensity vector) is to the right. 
  Detour paths (\reva{schematic} in \reva{pairs {\bf h}, {\bf i}, {\bf
  j}; see Figure~\ref{fig:detours}}) combine \reva{transitions
  involving} other levels and may include analogous bound--free
  transitions.
  The upper row shows pair combinations contributing line extinction:
  collisional photon destruction ({\bf a}), scattering out of the beam
  ({\bf b} and {\bf c}), photon conversion out of the beam ({\bf h},
  \revapar into other-wavelength photons and/or kinetic energy).
  The lower row shows pairs contributing line emissivity: collisional
  photon creation ({\bf d} and {\bf e}), scattering into the beam
  ({\bf f} and {\bf g}), detour photon production into the beam ({\bf
  i} and {\bf j}). 
  Pairs {\bf c} and {\bf g} have equal probability by requiring one
  photon in the beam and \reva{one} with arbitrary direction.
  From
  \href{http://www.staff.science.uu.nl/~rutte101/rrweb/rjr-edu/lectures/rutten_ssf_lec.pdf}{SSF},
  extension of \linkrtsa{85}{Figure~3.3}. 
  }
\label{fig:pairs}
\end{figure}

\begin{figure}
  \centering
  \includegraphics[width=45mm]{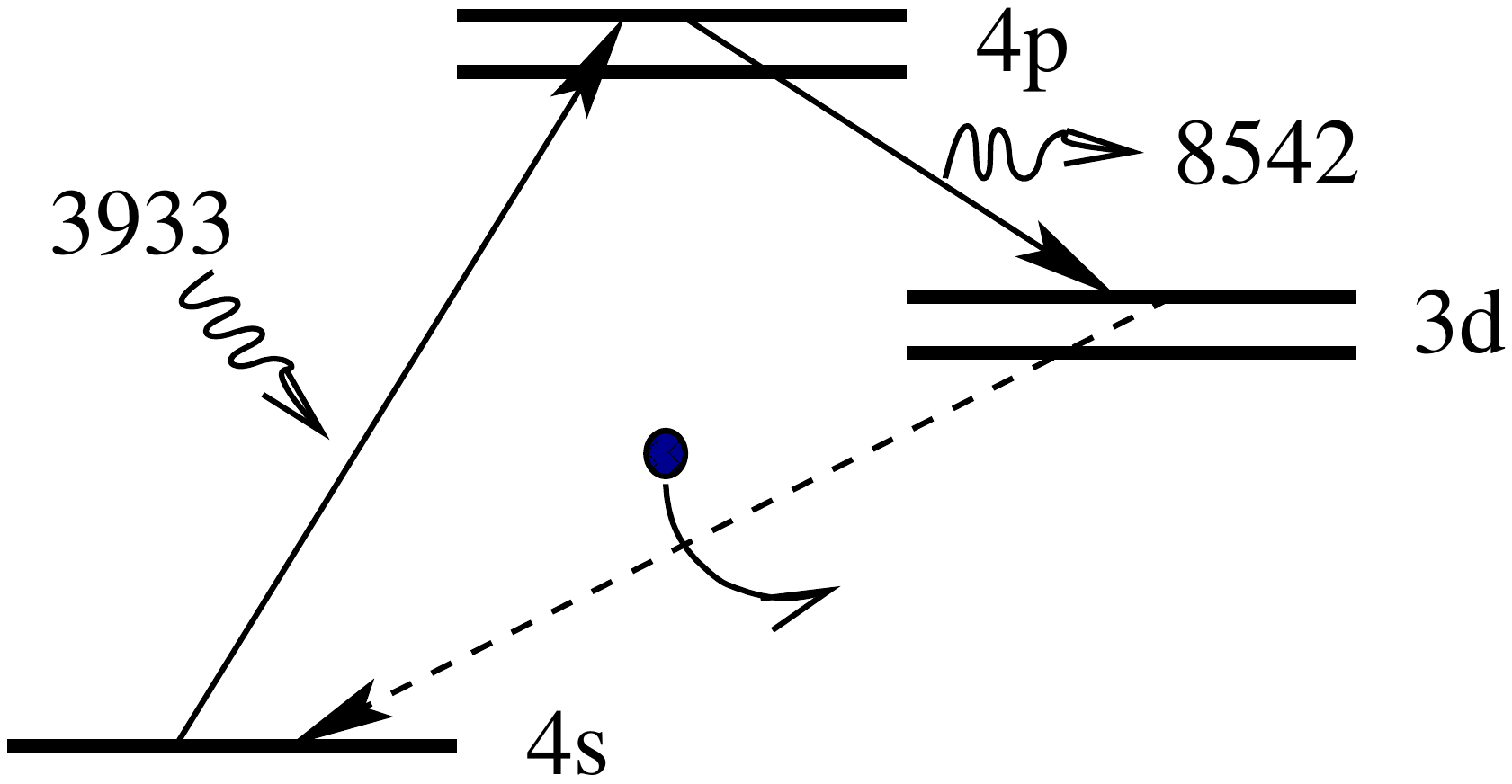}
  \hspace{20mm}
  \includegraphics[width=45mm]{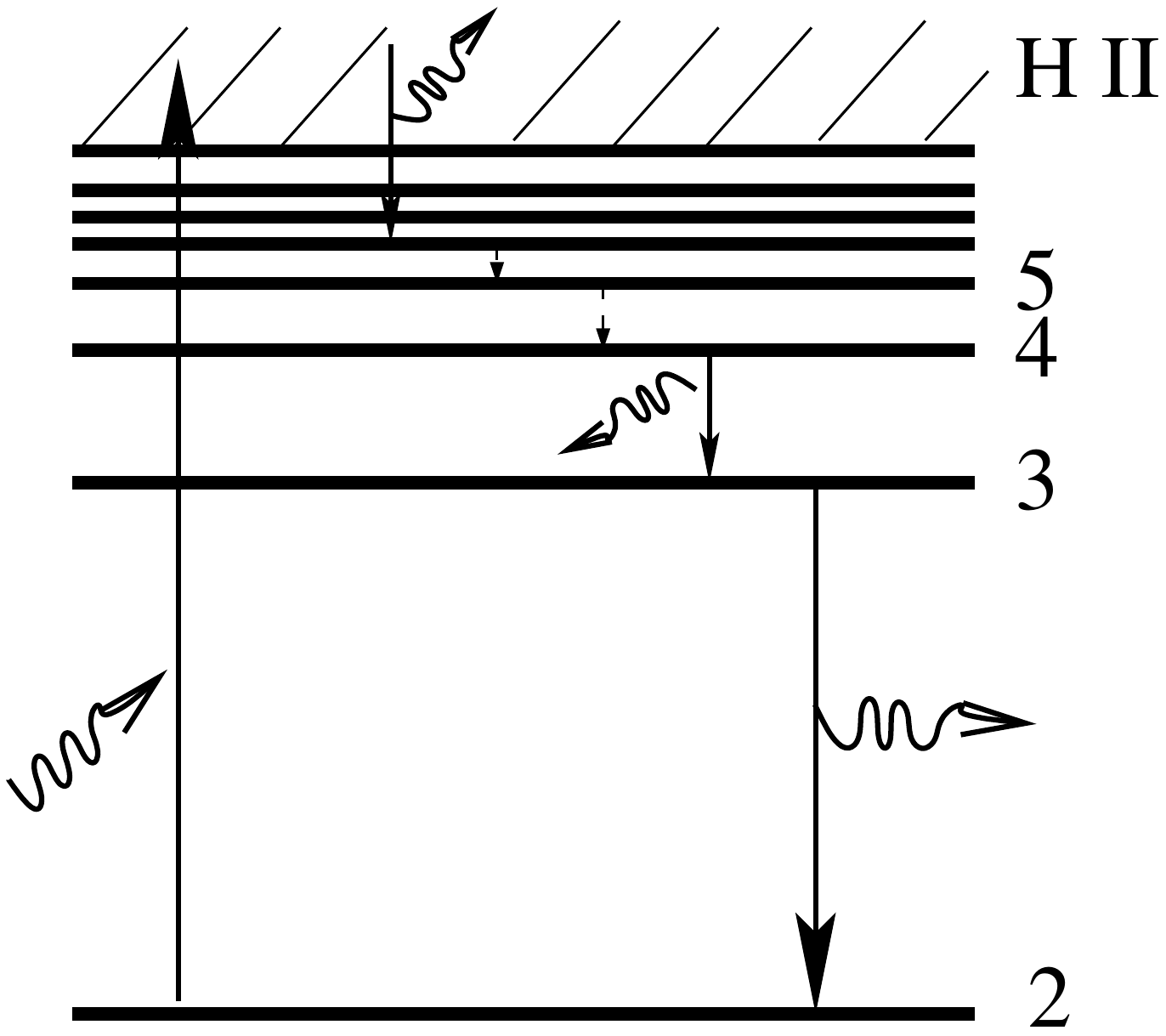}
  \caption[]{
  \reva{ Detour examples.
  {\em Left}: \CaIR\ emission in a bound-bound ``interlocking'' detour
  loop: $3d$--$4s$ down per collision (forbidden transition) followed
  by $4s$--$4p$ \CaIIK\ photo-excitation up and $4p$--$3d$
  photo-deexcitation down back to the $3d$ level, adding a \CaIR\
  photon to the beam.
  Starting at the $4s$ \CaII\ ground level the loop back to it
  extincts a \CaIIK\ photon by conversion into a \CaIR\ photon plus
  kinetic energy. 
  {\em Right}: \Halpha\ emission in a bound--free detour loop:
  photo-ionization from \HI\ 
  
  $n\tis\,2$ by the Balmer continuum
  followed by photo-recombination into a Rydberg level and downward
  cascade with $\Delta n\tis\,1$ steps, first as collisional
  deexcitations and then as photo-deexcitations ending by adding an
  \Halpha\ photon to the beam.
  The extincted Balmer-continuum photon is converted into
  longer-wavelength photons plus kinetic energy.
  }}
\label{fig:detours}
\end{figure}

\section{Equilibria} \label{sec:equilibria}

Figure~\ref{fig:pairs} is a pairwise inventory \reva{of the five
Einstein bound--bound processes affecting photons
(\linkrtsa{39}{Section~2.3.1}): photo-excitation (at left in pair {\bf
a}), spontaneous photo-deexcitation (at right in pair {\bf b}),
induced photo-deexcitation (in pair {\bf c}), collisional excitation
(in pair {\bf d}), and collisional deexcitation (in pair {\bf a}).}
The pairs at left describe two-level-only combinations; the pairs at
right multi-level detours \reva{where ``detour'' represents the sum of
all possible indirect transition paths from the upper level to the
lower level or {\em vice versa\/}; common cases are shown in
Figure~\ref{fig:detours}.}
\reva{The diagrams in Figure~\ref{fig:pairs}} can be drawn similarly
for bound--free ionization/recombination transitions.

Assuming statistical equilibrium (SE, constant level populations with
time) gives two extreme equilibria: \acp{LTE} when collisional pairs
{\bf a}, {\bf d}, and {\bf e} dominate, \revapar 
coronal equilibrium (CE) when pair {\bf d} dominates exclusively. 

\acp{LTE} requires sufficiently high density \revapar that most excited
atoms already deexcite per collision before doing so radiatively
(spontaneous or induced) within \reva{their excited-state
lifetime.} 
\reva{LTE} is valid throughout the Sun up to \reva{its} surface. 
Because \acp{LTE} requires colliders both up and down \reva{the
corresponding} Boltzmann upper-to-lower level, population ratios within
an ionization stage depend only on temperature.
\reva{Saha} upper-to-lower ion stage population ratios \reva{depend
additionally on} the electron density because, in addition to
\acp{LTE}-enforcing \reva{colliders both for ionization and
recombination, another} electron needs to be caught for recombination.  

\acp{CE} requires sufficiently low gas and radiation densities that
every collisional excitation (ionization) is followed by spontaneous
deexcitation (photorecombination). 
\reva{CE} is valid for most \acp{EUV} lines from the corona. 
\revapar The stage ratios \reva{then} depend only on temperature, the level
ratios also on collider (electron) density.

\acp{NLTE} describes \acp{SE} situations between these two extremes. 
The name is a misnomer, meaning \reva{assuming} \acp{SE} without
\reva{assuming} \acp{LTE}. 
\reva{It may encompass the above extremes: for example, LTE} is reached
at the bottom of the \acp{NLTE} VALIIIC atmosphere of
\citetads{1981ApJS...45..635V}, 
\revapar \acp{CE} at its top.
``\acp{NLTE} departures'' mean population differences with
Saha--Boltzmann values, source function differences with the Planck
function.

An additional \acp{NLTE} complexity is the issue whether in resonance
scattering (pairs {\bf f}, {\bf g}) the new beam photon ``remembers''
the precise wavelength of the exciting photon
(\linkadspage{1929MNRAS..89..620E}{2}{p.~2} of
\citeads{1929MNRAS..89..620E}). 
In coherent scattering it has the same wavelength, in complete
redistribution (CRD) it resamples the profile. 
Partial frequency redistribution (PRD) evaluates coherent scattering
with Doppler redistribution \revapar and collisional redistribution at
high collider density.
\reva{Doppler redistribution occurs always because} even when atoms
scatter coherently in their own frame the observer sees an ensemble
\revapar sampling different particle motions. 
In the case of systematic motions the line source function becomes
anisotropic and angle redistribution must also be accounted for.
For most solar lines \acp{CRD} is a sound assumption (generally made
since \citeads{1942PhDT........12H}) 
but \CaII\ \HK, \MgII\ \hk, \Lyalpha, and other strong ultraviolet
lines with high-up core formation at low density require \acp{PRD}
modeling.\footnote{Frequency and angle redistribution are not yet
treated in RTSA nor here.
\acp{ADS}-available \linkadspage{1968slf..book.....J}{110}{Chapter~5}
of \citetads{1968slf..book.....J} 
remains a good read (but read emissivity for emission coefficient).
The recent study by
\citetads{2017A&A...597A..46S} 
includes a good introduction and key references.}

Bound-free transitions are not intrinsically different from
bound--bound transitions (the rate descriptions can be unified, see
\linkrtsa{68}{Section~3.2.3}); they differ only in the larger extent and
threshold \reva{cutoff} of their profile function and in obeying
complete redistribution over that since the electron caught for
recombination samples the Maxwell distribution without memory.

Non-E is the \reva{non-SE} generalization of \acp{NLTE} to
time-dependent populations. \revapar \reva{Non-E can be} important in
situations where gas cools after being heated because the settling
\reva{speed} to collisional equilibrium has near-Boltzmann temperature
sensitivity \reva{from the Einstein relation between collisional up-
and down rates (\linkrtsa{70}{Eqs.~3.32 and 3.33}).}
Settling slow-down in cooling gas is most important in the large \HI\
\Lyalpha\ jump \reva{and causes} \revapar \reva{retarded} hydrogen
\reva{recombination} \revapar \reva{discussed further in
Section~\ref{sec:nonEmm}.
Non-E is likely also important in ionization/recombination settling
of} other species with high $n\tis\,2$ excitation energy
\reva{including \HeI\ }
(\citeads{2014ApJ...784...30G}), 
\revapar \reva{\SiIII\ and \OIII\
(\citeads{2018ApJ...858....8N})}. 

With ``non-equilibrium spectrum formation'' in the title I mean
relaxing \acp{LTE} into \acp{NLTE} for the formation of both lines and
continua, \reva{discarding the assumption of} static magnetism-free
hydrostatic equilibrium by using dynamical MHD simulations, and
\reva{replacing} \acp{SE} by non-E where necessary both in \revapar
simulations and in  \reva{spectral} synthesis.

\begin{figure}      
  \centering
  \includegraphics[width=8cm]{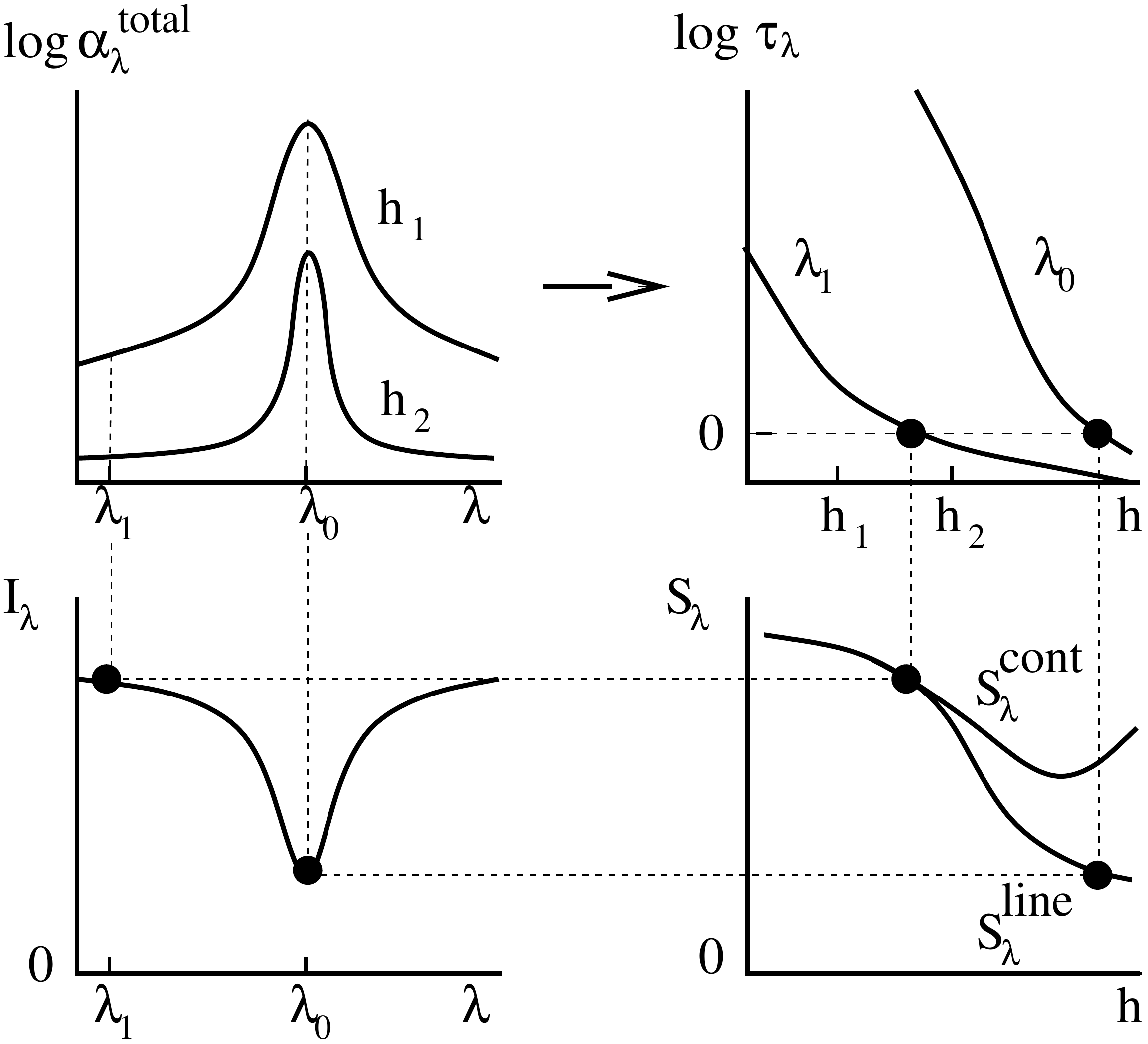}
  \caption[]{
  Schematic optically thick solar line formation.
  The line extinction $\alpha^l$ diminishes with height with \revapar
  density, \revapar \revc{likely} also with temperature (Boltzmann
  excitation), \reva{increasing} ionization, \revc{and} more.
  In the absence of systematic motions its extinction profile is
  symmetric around line center. 
  It has damping wings at lower height (h1) 
  from larger collider
  density, while its Doppler core narrows with height (from lower
  temperature \reva{and/}or \reva{less} \revapar microturbulence). 
  The total extinction ({\em upper left}) is its sum with the local
  continuum extinction $\alpha^c_\lambda$, which is typically orders
  of magnitude less \revb{(Figure~\ref{fig:extinction})}, hence the
  logarithmic scale.
  Summation of $\alpha_\lambda^{\rm total}$ against height delivers
  the optical depth scaling ({\em upper right}) which is usually
  roughly linear in $\log \tau(h)$ due to hydrostatic exponential
  density decay.
  The $\tau\tis\,1$ locations define where to sample the total source
  function $S_\lambda$ to obtain the emergent intensity in
  Eddington--Barbier fashion ({\em lower row}).
  $S_\lambda$ is the extinction-weighted combination of
  $S_\lambda^{\rm line}$ and $S_\lambda^{\rm cont}$.
  The emergent line
  $I_\lambda\tapprox\,S_\lambda(\tau_\lambda\tis\,1)$ is in absorption
  because $S_\lambda$ drops; for an optically thick emission line
  $S_\lambda$ must increase with height.
  This schematic mimics the formation of the solar \NaD\ lines in
  \linkadspage{1992A&A...265..268U}{5}{Figure~4} of
  \citetads{1992A&A...265..268U} 
  by having $S_\lambda^{\rm cont}\tapprox\,B_\lambda$ with a
  temperature minimum and a $\sqrt{\varepsilon}$-like scattering drop
  in $S_\lambda^{\rm line}$. 
  In \acp{LTE} the line would show a core reversal.  
  This quartet may be similarly drawn for bound--free transitions,
  with as major change the wide spectral extent and non-symmetric
  triangular shape of the extinction edge in the first panel. 
  In the \reva{lower-right} panel it is then more realistic to draw
  $S_\lambda^{\rm edge}$ as less steeply decaying than $B_\lambda$ and
  without outer rise, as in Figure~\ref{fig:schemscat}.
  From
  \href{http://www.staff.science.uu.nl/~rutte101/rrweb/rjr-edu/lectures/rutten_ssf_lec.pdf}{SSF}.
  }
\label{fig:4panel}
\end{figure}

\section{\reva{Spectrum Formation in a Nutshell}} \label{sec:equations}

\reva{This section is a brief summary of radiative transfer in stellar
atmospheres, with key equations.
It refers much to RTSA, with page openers.
The bibles of the field are
Mihalas 
(\citeyearads{1970stat.book.....M}, 
\citeyearads{1978stat.book.....M})  
and \citetads{2014tsa..book.....H}. 
Good summaries of numerical approaches are given by
\citetads{2003ASPC..288...31W} 
and \citetads{2019AdSpR..63.1434P}. 
}

Figure~\ref{fig:4panel} illustrates \revapar optically thick solar
spectrum formation \reva{using} the 
Edding\-ton--Barbier approximation\footnote{\revc{Nomenclature:} 
its first formulation was already in
\linkadspage{1921MNRAS..81..361M}{9}{Eq.~36} of
\citetads{1921MNRAS..81..361M} 
whereas its importance for interpreting spectral feature formation and
limb darkening was pointed out only much later by Uns\"old 
\reva{while} the hint in
\citetads{1926ics..book.....E} 
alluded to by \citetads{1943AnAp....6..113B} 
was indirect and unclear; 
see \citetads{2018OAst...27...76P}.} 
\begin{equation}
  I_\lambda(0,\mu) \approx S_\lambda(\tau_{\lambda\mu}\tis\,1) 
  \mbox{~~~or~~~}
    I_\lambda(0,\mu) \approx S_\lambda(\tau_\lambda\tis\,\mu) 
  \label{eq:EB}
\end{equation}
for the monochromatic emergent intensity $I_\lambda(0,\mu)$ in
direction $\mu \!\equiv\! \cos \theta$, with $\theta$ the viewing
angle between line of sight and local vertical and $S_\lambda$ the total
source function.
\reva{The optical depth $\tau_{\lambda\mu}$ is} measured inwards by
summing the extinction per cm $\alpha_\lambda$ along the line of sight
\reva{with $\tau\tis\,0$ the mathematical outer surface (in your
telescope)}, \reva{whereas the more common} $\tau_\lambda(h_0)
\!\equiv\! - \int_\infty^{h_0} \alpha_\lambda \dif h $ is the radial
optical depth for axial symmetry (plane-parallel layers).
\revapar

This relation is obtained by linearizing the Laplace transform of the
source function (\linkrtsa{106}{p.~86}), which is the solution of
the integral form of the intensity transfer equation in an outward
direction at the surface of a non-irradiated star
(\linkrtsa{97}{Eq.~4.10}).
Although this is only an approximation and can go
wrong\footnote{\href{http://www.staff.science.uu.nl/~rutte101/rrweb/rjr-edu/lectures/rutten_ssf_lec.pdf}{SSF} E-B exam: estimate the height of formation 
of the blend at 5889.76~\AA\ in the \NaI\ \Dtwo\ wing using
\linkadspage{1992A&A...265..268U}{5}{Figure~4} of
\citetads{1992A&A...265..268U}. 
Your Eddington--Barbier estimate of 150~km where the dip intensity
equals the value of $S\tapprox\,B$ falls a million times short -- the
line is telluric.}, it says that one should inspect the source
function where the summed extinction reaches unity, the observational
``surface''. 
The latter says where one looks, the former what one may see.
This recipe holds alike for solar observations, 1D modeling, and
numerical simulations. 

\reva{The recipe} does not hold for optically thin \reva{features (\eg\
filaments)} for which one instead quantifies total emissivity
$j_\lambda$ along the line of sight, requiring problematic accounting
for irradiation from below and sideways unless radiation-free \acp{CE}
can be assumed as \reva{is done} for coronal \acp{EUV} lines.

\reva{Thus, for photospheric and chromospheric radiation} \revapar one
\revapar evaluates the extinction for ``where'' and the source
function for ``what''.
These quantities are more orthogonal than extinction and emissivity:
to first order the extinction describes the local density of the
particular particles that contribute \reva{extinction} at a specific
wavelength whereas the source function describes the local environment
(gas density, temperature and impinging radiation) governing what
happens after extinction (photon destruction, scattering, or
conversion in Figure~\ref{fig:pairs}).
In \acp{LTE} this orthogonality is perfect: $\alpha_\lambda$ and
$j_\lambda$ \reva{then} share the same bound--bound spike while
$S_\lambda\tis\,B_\lambda$ \reva{with $B_\lambda$ the Planck function,
which} is smooth across a line ignoring its existence.
However, \reva{this} split \reva{between where and what} can get
mixed: for example in much of the solar atmosphere the extinction in
\Halpha\ \reva{(``where'')} is set by radiation in \Lyalpha\ with its
own ``what'' conditioning\footnote{E.g.\ around hot Ellerman bombs
that radiate \Lyalpha\ \reva{boosting \Halpha\ extinction} into
surrounding cool gas (\citeads{2016A&A...590A.124R}).}. 

Equation~\ref{eq:EB} is for total extinction and total source
function.
Continuous and line extinctions add up directly being cross-sections
($\alpha_\lambda$ per cm may be seen as volume coefficient with
cross-section cm$^2$ per cm$^3$), with the caveat that the emissivity
from induced deexcitations be counted as negative extinction to
accommodate the mutual cancelation\footnote{See my 
Richard N. Thomas memorial
\href{http://www.staff.science.uu.nl/~rutte101/rrweb/rjr-pubs/2003-thomas-epsilon.pdf}{``Epsilon''} \citep{Rutten2003a}.} 
of pairs {\bf c} and {\bf g} in Figure~\ref{fig:pairs}.
The total source function (emissivity divided by extinction; the word
{\em source\/} implies addition of new photons into the beam) is the
weighted mean over all contributing processes
$S_\lambda^{\rm tot}= \sum j_\lambda/\sum \alpha_\lambda$, in
particular
$S_\lambda^{\rm tot} = (j^c_\lambda + j^l_\lambda)/(\alpha^c_\lambda +
\alpha^l_\lambda) = (S_\lambda^c + \eta_\lambda S_\lambda^l)/(1 +
\eta_\lambda)$
where this $\eta_\lambda$ is the line strength
$\alpha_\lambda^l / \alpha_\lambda^c$ and $S_\lambda^{\rm tot}$ is
always frequency-dependent across a line even when $S_\lambda^l$ is
not by obeying \acp{CRD}.

Since both the line source function and the line extinction depend on
the lower- and upper-level population densities we employ shorthand
\acp{NLTE} departure coefficients (\linkrtsa{53}{Section~2.6.2}):
\begin{equation}
  b_l = n_l / n^{\rm LTE}_l
  \mbox{~~~~~~~~}
  b_u = n_u / n^{\rm LTE}_u,
\label{eq:zwaan}
\end{equation}
where $n^{\rm LTE}$ is the population density of the level computed
per Saha--Boltzmann from the total element density $N_{\rm
elem}$\footnote{\reva{Equation~\ref{eq:zwaan} uses} the Zwaan
definition of \citetads{1972SoPh...23..265W}. 
The Harvard definition following
\citetads{1937ApJ....85...88M} 
used by Avrett and Fontenla has $b^{\rm Menzel} \equiv n/n_c$
normalization by the next ionization stage; for majority-stage levels
containing most of the element $b^{\rm Menzel}\tapprox\,n_c^{\rm LTE} =
1/b_c^{\rm Zwaan}$ as carefully stated on
\linkadspage{1981ApJS...45..635V}{30}{p.~663} of
\citetads{1981ApJS...45..635V} but misinterpreted by Fontenla \etal\
(\citeyearads{2009ApJ...707..482F}), 
see \citetads{2012A&A...540A..86R}.}. 
The general line extinction coefficient so becomes
(\linkrtsa{54}{Eq.~2.111}, \linkrtsa{224}{Eq.~9.6}):
\begin{equation}
  \alpha^l_\lambda = \frac{\pi e^2}{m_\rme c} \,
  \frac{\lambda^2}{c}\, b_l \, \frac{n_l^{\rm LTE}}{N_{\rm elem}} \, N_\rmH \,
  A_{\rm elem} \, f_{lu} \, \varphi \left[1-\frac{b_u}{b_l} \,
  \frac{\chi}{\varphi}\, \ep{-hc/\lambda kT}\right],
  \label{eq:genalpha}
\end{equation}
where $A_{\rm elem}$ is the relative element abundance with
$N_{\rm elem}\tis\,A_{\rm elem} N_\rmH$, $f_{lu}$ is the oscillator
strength, and \revapar $\chi$ and $\varphi$ are the area-normalized
profile functions for \revapar induced emission and
extinction.

The general line source function is (\linkrtsa{53}{Eq.~2.105}):
\begin{equation}
   S_\lambda^l =  \frac{2 h c^2}{\lambda^5} \frac{\psi/\varphi}
                   {\displaystyle \frac{b_l}{b_u} \ep{h c/\lambda k T}
                                                -\frac{\chi}{\varphi}}
  \label{eq:genS}
\end{equation}
\reva{with $\psi$ the spontaneous emission profile}. 
For \acp{CRD} $\psi = \chi = \varphi$ so that the profile ratios
simplify to unity. 
In the Wien approximation, generally valid throughout the
optical and ultraviolet (\Halpha\ reaches $\lambda T\tis\,hc/k$ at
21\,900\,K), the \acp{CRD} expressions simplify further to:
\begin{eqnarray}
   \alpha_\lambda^l & \approx & b_l \,\, \alpha_\lambda ^{\rm LTE} 
        \label{eq:Wienalpha}\\
   S_\lambda ^l & \approx & (b_u/b_l)\,\, B_\lambda(T),
        \label{eq:WienS}
\end{eqnarray}
which are the quick NLTE recipes for ``where'' and ``what'':
line extinction scales with $b_l$, the line source function with
$b_u/b_l$.
The latter gives $S_\lambda^l\tis\,B_\lambda$ for $b_u\tis\,b_l$; this
is not the definition of \acp{LTE} but a corollary: \acp{LTE} is
defined as Saha--Boltzmann partitioning with
$b_u\tis\,b_l\tis\,1$ 
(Section~1.4 of \citeads{1973trsl.book.....I}). 

Equation~\ref{eq:WienS} quantifies \acp{NLTE} departure from the Planck
function but not how it comes about. 
This needs splitting the line extinction coefficient into the 
destruction (a for absorption), scattering (s), and detour (d)
contributions of Figure~\ref{fig:pairs}:
\begin{equation}
  \alpha_\lambda^l \equiv  
    \alpha_\lambda^\rma +\alpha_\lambda^\rms + \alpha_\lambda^\rmd
  \hspace{10mm}
  \varepsilon_\lambda \equiv \alpha_\lambda^\rma / \alpha_\lambda^l
  \hspace{10mm}
   \eta_\lambda \equiv \alpha_\lambda^\rmd / \alpha_\lambda^l 
   \label{eq:epseta} 
\end{equation}
where $\varepsilon$ is the collisional destruction probability of an
extincted photon and $\eta$ is its detour conversion probability.
With these the general line source 
function\footnote{The multilevel detour terms are not yet 
added to \linkrtsa{84}{Section~3.4}.
\revapar 
The best description remains 
\linkadspage{1968slf..book.....J}{199}{Section~8.1} 
of \citetads{1968slf..book.....J} 
(with $\varepsilon \equiv \alpha^\rma/\alpha^\rms$ and 
$\eta \equiv \alpha^\rmd/\alpha^\rms$; 
remove the minus in the 
\linkadspage{1968slf..book.....J}{200}{equation after 8.8}).}
becomes \reva{either}
\begin{equation}
 S_\lambda^l = (1 - \varepsilon_\lambda - \eta_\lambda) \, J_\lambda 
  ~+~ \varepsilon_\lambda B_\lambda(T) ~+~ \eta_\lambda S_\lambda^\rmd 
\label{eq:S_CS} 
\end{equation}
\reva{or}
\begin{equation}
 S_{\lambda_0}^l =  (1 - \varepsilon_{\lambda_0} 
     - \eta_{\lambda_0}) \, \overline{J_{\lambda_0}} 
     ~+~ \varepsilon_{\lambda_0} B_{\lambda_0}(T) 
     ~+~ \eta_{\lambda_0} S_{\lambda_0}^\rmd.
\label{eq:S_CRD}
\end{equation}
The first version is for coherent (monofrequent, monochromatic)
scattering.
The  second is for \acp{CRD} with
$\overline{J_{\lambda_0}} \equiv (1/4\pi)\int\!\!\int I_\lambda
\varphi(\lambda\!-\!\lambda_0) \dif\lambda \dif\Omega$
the ``mean mean'' intensity averaged over all directions and the line
profile, with $\lambda_0$ the line-center wavelength and also used as
line identifier.  
The first term in Eqs.~\ref{eq:S_CS} and \ref{eq:S_CRD} represents the
reservoir of photons that contribute scattering, the second describes
collisional photon creation, the third the contribution of new line
photons via detours. 
The \acp{LTE} $S\tis\,B$ equality holds when
\reva{$\varepsilon\tis\,1, \eta\tis\,0$} and/or
\reva{$J\tis\,S^\rmd\tis\,B$}, both true below
the standard $h\tis\,0$ surface at $\tau_{5000}^c\tis\,1$. 
Above it $\varepsilon$ becomes small from lower electron density while
$\eta$ is usually smaller; all bound--bound lines and bound--free
continua formed above a few hundred km height are heavily scattering
with $S\tapprox\,J$ (but free--free continua always have $S\tis\,B$
because each interaction is collisional).  

\Halpha\ has a sizable $\eta S^\rmd$ contribution \reva{from detour loops
as in Figure~\ref{fig:detours}.  \revd{This} was} famously called
``photoelectric control'' by \citetads{1957ApJ...125..260T} 
\revd{but} wrongly because even this \reva{complex} \revapar 
line is mostly scattering, \revd{although} with unusual backscattering from
chromosphere to photosphere
(\citeads{2012A&A...540A..86R}). 

The final \reva{and most important} equation is the Schwarzschild
equation (\reva{page 361 of
\citeads{2014tsa..book.....H};} 
\linkrtsa{98}{Eq.~4.14}):
\begin{equation}
  J_\lambda(\tau_\lambda) = 
  \frac{1}{2} \int_0^\infty S_\lambda(t_\lambda) \, E_1(|t_\lambda
  \!-\! \tau_\lambda|) \dif t_\lambda 
  \equiv \lambdop_{\tau_\lambda}[S_\lambda(t_\lambda)]
\label{eq:lambdop}
\end{equation}
with exponential integral
$E_1(x)=\int_0^1 \ep{-x/\mu} \, \rmd \mu / \mu$
(\linkrtsa{98}{Eq.~4.12}).
Its weighting of the source function makes the kernel of the \lambdop\
operator very wide. 
The cutoff at the surface produces outward
$J_\lambda(\tau_\lambda) < S_\lambda(\tau_\lambda)$ divergence for
small inward increase of $S_\lambda(\tau_\lambda)$ but outward
$J_\lambda(\tau_\lambda) > S_\lambda(\tau_\lambda)$ divergence for
steep increase (\linkrtsa{100}{Figure~4.2}, \linkrtsa{103}{Figure~4.4}
from \citeads{1952bmtp.book.....K}, 
\linkrtsa{121}{Figure~4.9}).
Be aware that steep horizontal gradients are similarly important in 3D
radiative transfer as the radial ones entering \lambdop\ in
plane-parallel Eq.~\ref{eq:lambdop}.
\revapar
\reva{Steep horizontal gradients occur in and above granulation and
also in and around the MCs that make up network and plage and 
so} affect their irradiance contributions.

Equation~\ref{eq:lambdop} defined the industry of
``approximate/accelerated lambda iteration''
(\linkrtsa{145}{Section~5.3.2}) following the operator splitting of
\citetads{1973ApJ...185..621C}. 
The need for iteration when simplistic \acp{LTE} can not be assumed is
obvious from comparing Eqs.~\ref{eq:S_CRD} and \ref{eq:lambdop}:
determining $S$ for using Eq.~\ref{eq:EB} needs $J$ and that requires
$S$ over a range of depths.
\reva{Thus,} the observed intensity depends non-locally on radiation
from elsewhere, \reva{not} only for the source function for ``what''
but also for the extinction for ``where'' \reva{when} that senses
radiation (as in the extinction of \Halpha\ controlled by radiation in
\Lyalpha).
With overlapping transitions (the local continuum for starters) and
interlocking (the multi-level $\eta$-term) this involves all pertinent
transitions connected one way or other to the one of interest. 
This quickly grows into having to solve radiative transfer and
associated atomic-level population equations for many species at many
wavelengths in multiple directions throughout the atmosphere. 
Solar spectrum modeling so developed into a computational
resource-limited endeavor with clever code development of principal
importance. 

At its very start
\citetads{1965SAOSR.174..101A} 
computed his canonical demonstration (\linkrtsa{128}{Figure~4.12}) of
the quintessential $\sqrt{\varepsilon}$ law: in a two-level-atom
isothermal atmosphere with constant $\varepsilon$ the line source
function drops outward to $S^l = \sqrt{\varepsilon}\,B$ at the $\tau\tis0$
physical surface, both for coherent scattering and \acp{CRD}. 
The emergent intensity ($\tapprox\,S$ at the $\tau\tis\,1$ observed
surface) is nearly as small. 
The thermalization depth where $J$ starts dropping \reva{below} $B$ due
to scattering photon losses lies as deep as $1/\varepsilon$ in
line-center optical depth units, even deeper in the presence of
damping wings permitting \reva{long} photon steps 
(\linkrtsa{112}{Section~4.3} and
\href{https://www.cfa.harvard.edu/~avrett/nonltenotes.pdf}{Avrett's
lecture notes}).
\reva{The upshot is that scattering lines get very dark, as the \NaD\
lines in Figure~\ref{fig:haze} with their formation sketched in
Figure~\ref{fig:4panel}.}    
A simple explanation is given in Section~1.7 of
\citetads{1986rpa..book.....R}, 
an elaborate one in \citetads{1987A&A...185..332H}. 

\begin{figure}      
  \centering
  \includegraphics[height=30mm]{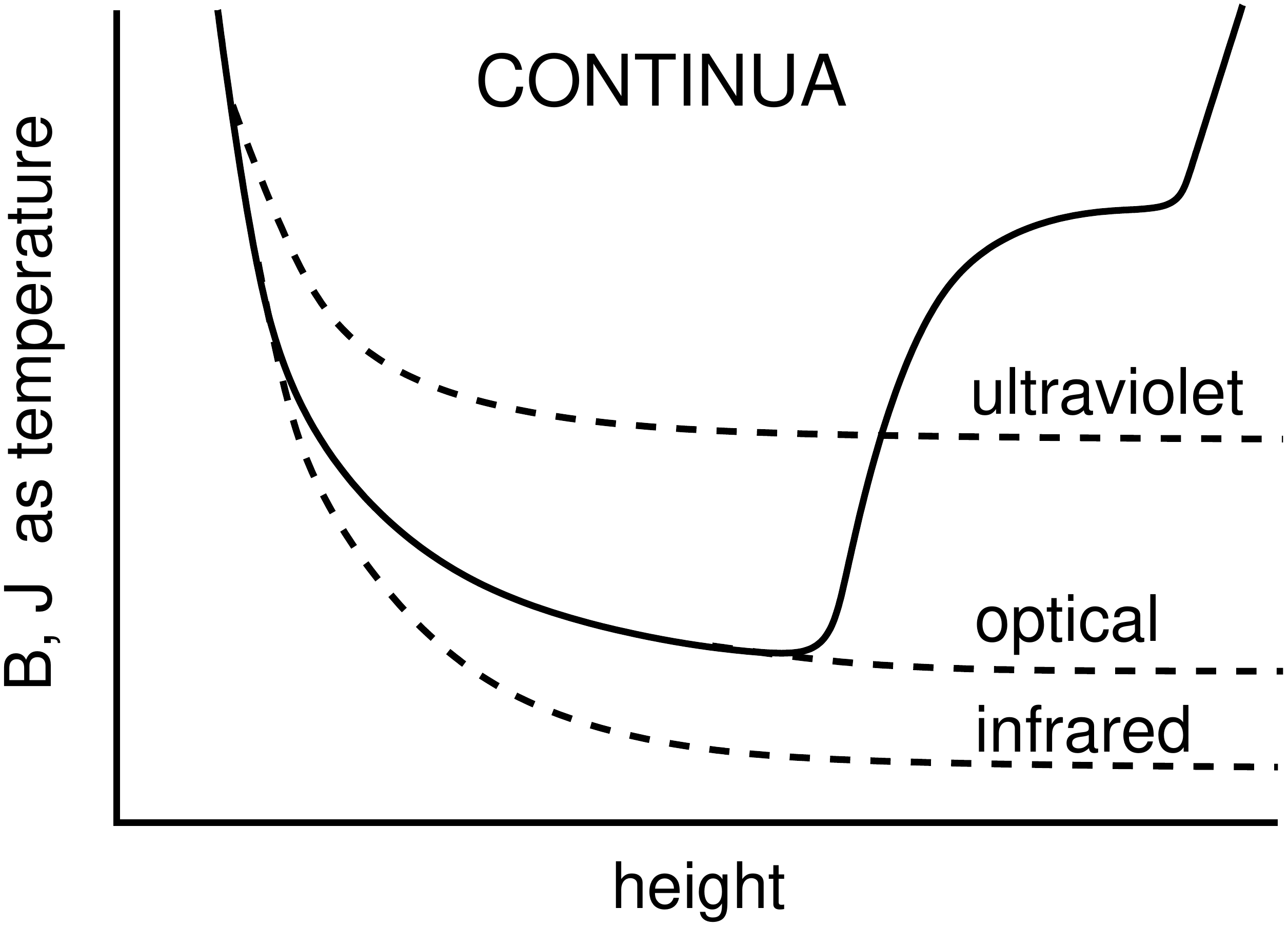}
  \hspace{5mm}
  \includegraphics[height=30mm]{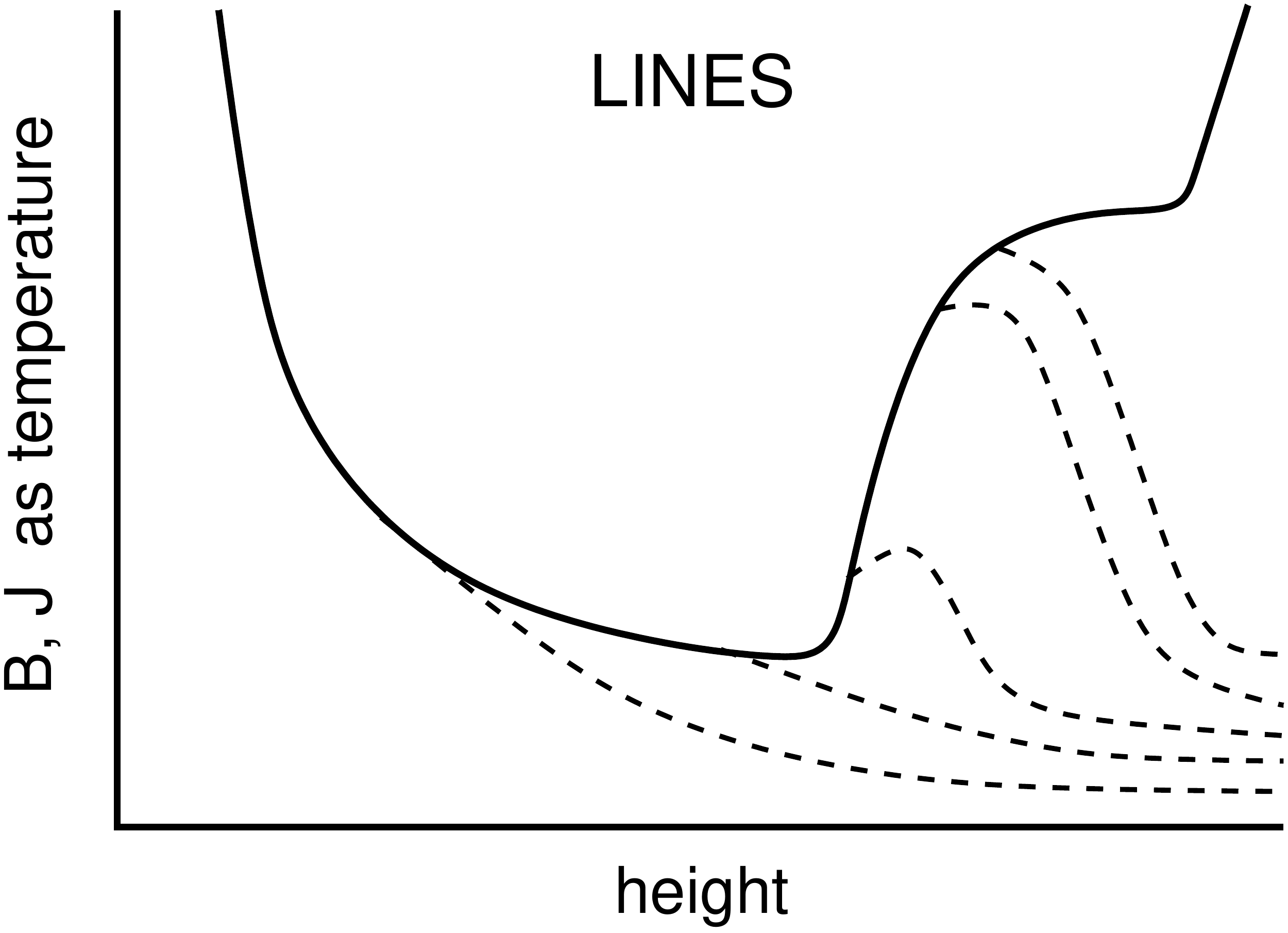}
  \caption[]{
  Scattering in the solar atmosphere. 
  {\em Left}: continua.
  {\em Right}: lines. 
  $B$ \reva{({\em solid\/})} and $J$ \reva{({\em dashed\/})} are shown
  as temperature to have the same scales at different wavelengths.
  \revapar The $B$ \reva{curve} \revapar mimics the solar atmosphere
  in having a radiative-equilibrium decline, a higher-temperature
  chromosphere, and a steep increase to coronal values.
  From
  \href{http://www.staff.science.uu.nl/~rutte101/rrweb/rjr-edu/lectures/rutten_ssf_lec.pdf}{SSF}.
  }
\label{fig:schemscat}
\end{figure}

\section{\revb{Scattering Overview}} \label{sec:schemscat}
\reva{Figure~\ref{fig:schemscat} sketches the behavior of $B$ and $J$ in
continua and lines as electron temperature and radiation temperature.
Temperature representation (with excitation temperature for $S$, see
\linkrtsa{57}{p.~37\,ff.}) enables direct comparison between different
wavelengths by undoing Planck-function temperature sensitivities.}

The \reva{first sketch in} Figure~\ref{fig:schemscat} shows \revapar
\reva{continua with} $J > B$ \reva{in the ultraviolet} \reva{but} $J
\tapprox\,B$ in the optical and $J<B$ in the infrared.
The physical reason is that the upper photosphere (above the
granulation and below the heights where acoustic waves shock and
\acp{MC}s expand into canopies, still predominantly neutral) is the
most homogeneous domain of the solar atmosphere, fine-structured
primarily by \reva{non-shocking} acoustic and gravity waves, and
generally close to radiative equilibrium.
This condition requires a temperature decline producing
spectrum-integrated $\alpha S\tapprox\,\alpha J$
(\linkrtsa{173}{Section~7.3.2}).
\revapar {The bulk} of the solar radiation escapes in the optical and
\reva{so imposes} $S\tapprox\,J$ \reva{in this part of the spectrum.
The optical continuum has} $S\tapprox\,B$ thanks to \Hmin\
\reva{dominance, so that its radiative equilibrium} sets the
upper-photosphere temperature decline.

With this imposed \revapar decline \lambdop\ produces \revapar $J>B$
\reva{divergence in the ultraviolet.
This holds} already in \acp{LTE} from \reva{the} nonlinear Wien
\reva{sensitivity} (\linkrtsa{121}{Figure~4.9}): \reva{the optical $S
\tapprox\,B$ transforms into $J>B$ using \lambdop\ in the ultraviolet
because $B(h)$ is much steeper there.
Yet} larger $J>B$ divergence \reva{results} from \acp{NLTE} bound--free
scattering \reva{because the effective photon escape depths then lie} deeper
than $\tau\tapprox\,1$, \revapar sampling \reva{yet steeper temperature
increase} in the deep photosphere.
\revapar \reva{The ultraviolet photons that are collisionally created
there} scatter outward \reva{gaining $J > B$ because the actual $B(h)$
gradient is steeper than what they would themselves impose for
equilibrium.
Eventually} $J$ \reva{flattens} to constancy (out to infinite height
above infinite-extent plane-parallel models). 
In the chromosphere the independent non-equilibrium temperature rise
results in $J<B$ \reva{above this limit value}.
\reva{These ultraviolet $J \neq B$ patterns produce corresponding $S
\neq B$ departures where scattering dominates.}

The infrared and mm regions have $J<B$ from \lambdop\
(\linkrtsa{121}{Figure~4.9}) but this has no effect on $S$ since $S\tis
B$ for the scatter-free \Hmin\ and \HI\ free--free \reva{processes}.

The right-\reva{hand sketch in Figure~\ref{fig:schemscat}} depicts
scattering lines \reva{with} increasing \revapar extinction. 
The decoupling of $J$ from $B$ occurs higher for stronger lines. 
$J$ drops below $S$ qualitatively following the isothermal
$\sqrt{\varepsilon}$ law, also in the radiative-equilibrium decline
because at large line extinction the $\tau$ scale gets compressed so
that $B(\tau)$ drops less steeply than in the continuum
(\linkrtsa{123}{Figure~4.10}).

This \reva{sketch} also represents a schematic for the formation of
\acp{PRD} lines. 
Their core, inner-wing, and outer-wing parts \reva{represent}
independent radiation ensembles, each scattering out on its own
\reva{with its own decoupling height and decay from $B$}.
\reva{Across a strong line} the rightmost $J$ curve \revapar describes
\revapar line-center scattering, the ones \reva{more} to the left
scattering further away from line center with deeper decoupling.

\begin{figure}      
  \centering
  \includegraphics[width=8cm]{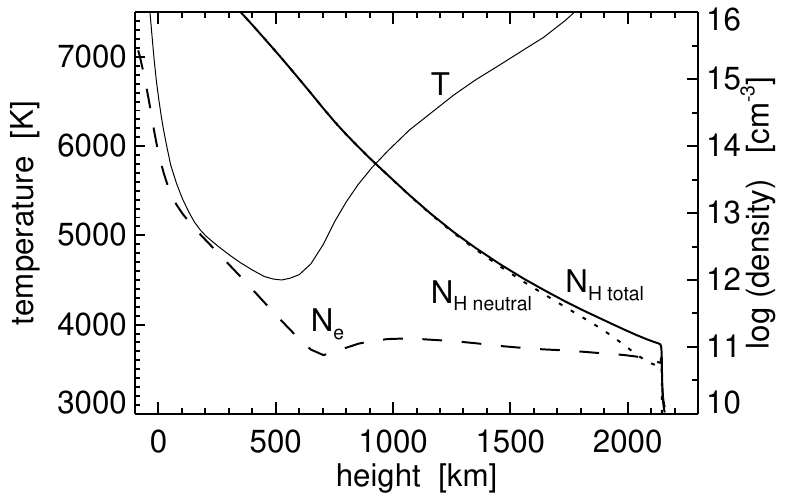}
  \caption[]{
  The FALC quiet-Sun model of
  \citetads{1993ApJ...406..319F}\revb{.} 
  The electron density $N_\rme$ \reva{({\em dashed})} has $10^{-4}$
  offset from the \reva{hydrostatically decaying total hydrogen
  density} $N_\rmH$ \reva{({\em solid\/})} over $h\tis\,100-700$\,km. 
  Below and above \reva{this range hydrogen has partial ionization
  above $10^{-4}$}, reaching full ionization \reva{below 0\,km and}
  \revapar above 2100\,km. 
  Above 1500\,km ionization makes the neutral hydrogen density
  $N_{\rmH\rmI}$ ({\em dotted}) drop below $N_\rmH$.
  \reva{After}
  \href{http://www.staff.science.uu.nl/~rutte101/rrweb/rjr-edu/lectures/rutten_ssx_lec.pdf}{SSX}.
  }
\label{fig:FALC}
\end{figure}

\section{FALC Demonstrations} \label{sec:demonstrations}

\reva{This section illustrates solar spectrum formation using examples
for a standard 1D model atmosphere to illustrate NLTE effects that
affect solar irradiance.}
The graphs \revapar were made \reva{with the \textsf{RH} spectral
synthesis code of \citetads{2001ApJ...557..389U} 
for the FALC model of
\citetads{1993ApJ...406..319F}}. 

\textsf{RH}
\revapar
does not iterate \lambdop\ but the emissivity $\Psi$ operator of
\citetads{1992A&A...262..209R}. 
It \reva{permits overlapping lines}, it includes \acp{PRD} \revc{and
full-Stokes options}, \revcpar it exists in 1D, 2D, 3D, spherical, and
Cartesian versions, and more recently also in parallel multi-column
``1.5D'' (\citeads{2015A&A...574A...3P}). 
Here \reva{its} 1D \reva{version-2} is used \reva{with H, He, Si, Al,
Mg, Fe, Ca, Na, and \revb{Ba} active, C, N, O, S, and Ni passive,} and
with 20\,m\AA\ sampling \revb{of 343\,000 lines between 1000 and
8000\,\AA\ in} the atomic and molecular line list of
\citetads{2009AIPC.1171...43K}. 

FALC, shown in Figure~\ref{fig:FALC},\footnote{\revc{The
\textsf{RH}-based stratifications in Figure~\ref{fig:FALC} differ
slightly (negligibly \revd{here}) at chromospheric heights from the
ones in \linkadspage{1993ApJ...406..319F}{7}{Table~2} of
\citetads{1993ApJ...406..319F} 
\revd{from mass-scale determination without ambipolar diffusion and
evaluation of electron densities with \textsf{RH}'s 
element mix (H.~Uitenbroek, private communication)}.}}
is the \revd{average-}quiet-Sun companion of the FALP plage model at
the basis of the \textsf{SATIRE} irradiance modeling. 
It is therefore used here rather than Avrett's latest quiet-Sun model
(ALC7 of \citeads{2008ApJS..175..229A}). 
Their differences are compared in
\href{http://www.staff.science.uu.nl/~rutte101/rrweb/rjr-edu/lectures/rutten_ssx_lec.pdf}{SSX}
but \revc{are} not significant for the demonstrations here.

While \revd{well-known and} often-cited, these \revc{1D} modeling
efforts \reva{do not describe the actual solar atmosphere
realistically}.
\revapar Even if its fine structures were just temperature
fluctuations around a well-defined mean temperature stratification,
\reva{then} the latter is not \reva{retrieved} by fitting the
mean-intensity spectrum in the optical and ultraviolet due to \revapar
non-linear Wien weighting, as shown for acoustic shocks by
\citetads{1995ApJ...440L..29C} 
and for granulation by
\citetads{2011ApJ...736...69U}. 
\revapar The 1D models should be regarded as hypothetical
plane-parallel stars with a spectrum remarkably mimicking the solar
spectrum \revapar \reva{that are} useful to demonstrate spectrum
formation governed by the equations above (\linkrtsa{209}{p.~189}). 
In RTSA I added didactic demonstrations from the monumental VALIII
modeling of \citetads{1981ApJS...45..635V} 
and its \reva{informative graphs} (\eg\
\linkrtsa{204}{Figures\,8.9\,ff.} from the 11-page
\linkadspage{1981ApJS...45..635V}{33}{Figure~36}).  
Here I add demonstrations with FALC and \textsf{RH}-output plotting
programs on
\href{http://www.staff.science.uu.nl/~rutte101/rridl/rhlib}{my
website}.

\begin{figure}
  \centering
  \includegraphics[width=10cm]{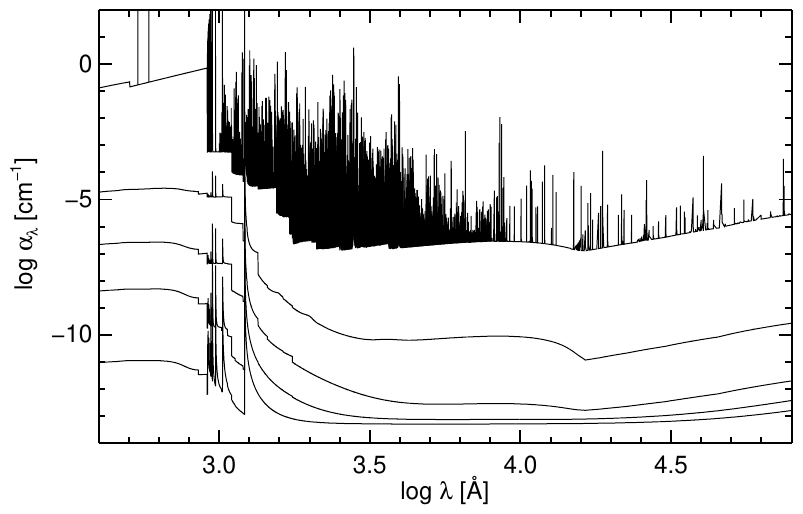}
  \caption[]{
  \revb{Extinction in the FALC star} \reva{\revbpar at heights 0, 500,
  1000, 1500, and 2000\,km from top to bottom.
  The top curve for $h\tis\,0$\,km is for total extinction: continuous
  plus all lines in the \textsf{RH} setup from active and passive
  atoms and the Kurucz-list sampling over 1000\,--\,8000\,\AA.
  The other curves are for continuous extinction only.
  The line haze \revb{is densest in the ultraviolet and violet}.
  At larger height the \revbpar ultraviolet edges are replaced by
  Rayleigh scattering \revb{off neutral and excited hydrogen atoms
  with $\propto\! \lambda^{-4}$ decreases}, the \revb{optical}
  \Hmin\,bf bulge below 1.6\,$\mu$m 
  ($\log\,\lambda[{\rm \AA}]\tis\,4.2$) by Thomson scattering
  \revb{off free electrons without wavelength variation}, and the
  \Hmin\,ff increase above 1.6\,$\mu$m by \HI\,ff extinction \revb{of
  protons with $\propto\! \lambda^2$ increase}.
  \revb{At $\log \alpha_\lambda\tis\,-7$ a 100-km thick feature
  becomes optically thick (\cf\ Figure~\ref{fig:extsb}).}
  Towards \reva{shorter and longer wavelengths} one observes higher in
  the FALC atmosphere, with $\tau_\lambda\tis\,1$ sampling of the
  chromosphere \reva{reached} below $\lambda\tapprox\,1600$\,\AA\
  ($\log\,\lambda[{\rm \AA}]\tis\,3.2$) and above
  $\lambda\tapprox\,160\,\mu$m  ($\log\,\lambda[{\rm \AA}]\tis\,6.2$)
  \revb{(\cf\ \linkadspage{1990IAUS..138....3A}{2}{Figure~1} of
  \citeads{1990IAUS..138....3A}). 
  }}} 
\label{fig:extinction}
\end{figure}

\reva{\subsection{FALC Extinction}} \label{sec:extinction}

\revapar \reva{Figure~\ref{fig:extinction} is an overview of FALC
extinction.}
\revb{Because FALC is a solar-like} high-metallicity star, ionization
of the electron donors (abundant elements with low first ionization
threshold: 
\SiI, \FeI, \AlI, and \MgI, see \linkrtsa{164}{Figure~7.1})
produces $N_\rme\tapprox\,10^{-4}\,N_\rmH$ throughout the photosphere,
enough to make it opaque\footnote{At much lower density 
than the transparent air around us.}
through combining with H into \Hmin.
\reva{At mm wavelengths} \revapar \Hmin\ free--free \reva{extinction
is} replaced by \HI\ free--free \reva{extinction} when the \revapar
$\tau_\lambda\tis\,1$ height reaches hydrogen
ionization\footnote{\revc{Nomenclature:}
\citetads{1868RSPS...17..131L} 
defined the chromosphere as an off-limb envelope radiating mostly in \HI\
Balmer lines and \HeIDthree, implying that it consists on-disk of what
is observed in \Halpha: a mass of fibrilar features constituting
the wildest 
scene in solar imaging. 
I regard these as product of \revc{small-scale} dynamic hydrogen
ionization (with partial field mapping from partial ionization and
retarded recombination)
and define the chromosphere as the solar-atmosphere regime where
hydrogen ionization reigns.}.
These free--free processes strictly obey source function
$S\tis\,B$ equality.
The \Hmin\ bound--free part governing the optical continua does not
share this virtue, but because the \Hmin\ ionization energy is less
than the average kinetic energy $S\tapprox\,B$ is
usually valid \reva{up to the heights were Thomson scattering takes over}.
In contrast, the \reva{Balmer and metal edges that together supply
increasing extinction} below \reva{3700}\,\AA\ are heavily scattering. 
Since the \reva{neutral metals} producing them also produce most lines in the
ultraviolet and optical (in particular \FeI), these are all affected
by this bound--free scattering as shown below.

\begin{figure}      
  \centering
  \includegraphics[height=38mm]{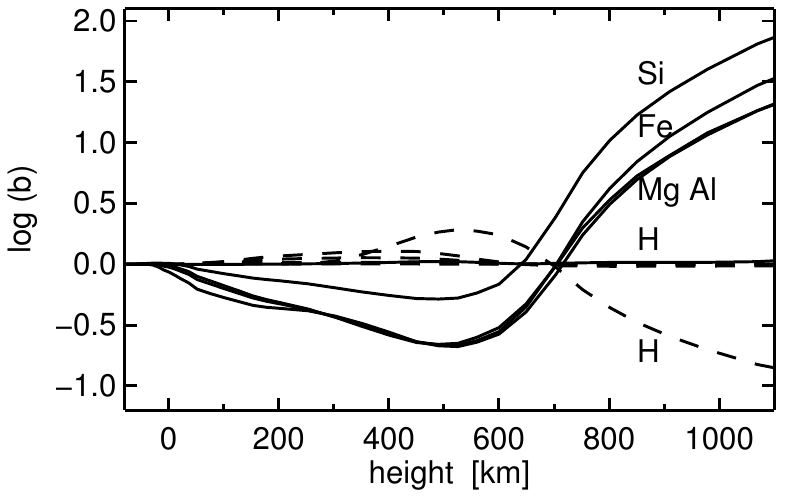}
  \includegraphics[height=38mm]{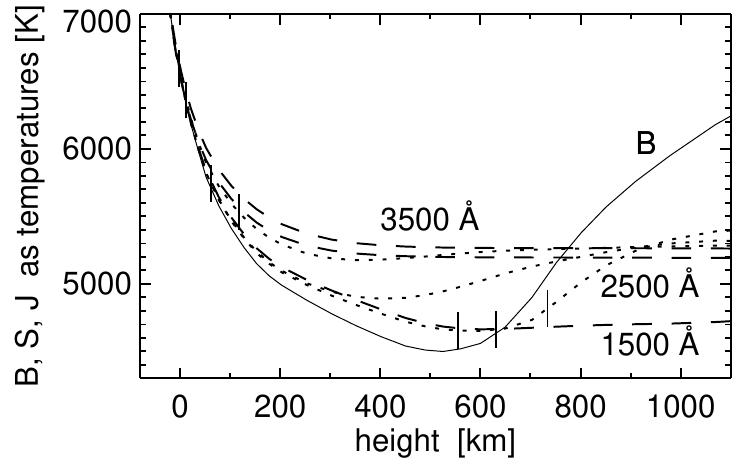}
  \caption[]{
  Ultraviolet continua in the FALC star. 
  {\em Left}: departure coefficients $b_1$ and $b_c$ for the neutral
  ({\em solid\/}) and ion \reva{({\em dashed\/}, all near unity)}
  ground states of the \reva{major} bound--free opacity providers Si,
  Fe, Al, and Mg.
  For hydrogen the solid line near unity shows $b_2$ for
  $n\tis\,2$, the dashed curve $b_c$ for the proton
  density.
  {\em Right}: $B$ ({\em solid\/}), $S$ ({\em dotted\/}), $J$ ({\em
  dashed}) averaged over 100\,\AA\ wide bands around the specified
  wavelengths including all Kurucz lines, for the atomic ones using
  the two-level scattering approximation (Section~\ref{sec:linehaze}).
  $S$, $B$ and $J$ are plotted as formal temperatures to obtain equal
  scales at different wavelengths.
  The ticks on \reva{each} $S$ \reva{curve} are at
  $\tau\tis\,3, 1, 0.3$ for 1500\,\AA, for $\tau\tis\,1$ and 0.3
  for 3500\,\AA\ and 2500\,\AA\ (the latter around $h\tis\,100$~km) for
  which the $\tau\tis\,3$ values lie below $h\tis\,0$~km near 7500\,K
  outside the frame. 
  \revc{At 2500\,\AA\ the latter lies so deep due to \MgI\ ionization
  (\linkadspage{1981ApJS...45..635V}{37}{VALIII Figure~36}).}
  After
  \href{http://www.staff.science.uu.nl/~rutte101/rrweb/rjr-edu/lectures/rutten_ssx_lec.pdf}{SSX}.
  }
\label{fig:edges}
\end{figure}

\subsection{FALC Ultraviolet Continua}  \label{sec:uvcontinua}

The ultraviolet continua from the Balmer threshold at 3646\,\AA\ down
to the Lyman threshold at 912\,\AA\ form at increasing height by the
summed extinction of the bound--free edges of \MgI, \AlI, \SiI, \FeI,
and \CI\ superposed on the Balmer edge (Figure~\ref{fig:extinction},
specification in \linkadspage{1981ApJS...45..635V}{32}{VALIII
Table~9}).

Figure~\ref{fig:edges} samples their FALC formation.
\reva{At left it shows representative population departures, at right
continuum formation at three representative wavelengths}.
\reva{The $\tau$ marks show that} at 3500\,\AA\ the continuum has
final photon escape \revapar in the deep FALC photosphere, at
2500\,\AA\ in the low photosphere, at 1500\,\AA\ in the onset of the
FALC temperature rise. 
However, the effective onset of scattering radiation escape
(thermalization depth) is \revapar in the deep photosphere for all
three continua \reva{as shown by their $J>B$ divergence}.
\revc{Assuming $S\tis\,B$ as in
\textsf{SATIRE}\footnote{\revc{\textsf{SATIRE} does not use FALC for
\revd{average-}quiet-Sun, but a Kurucz radiative-equilibrium model that is
closely the same in the photosphere
(\linkadspage{2012A&A...540A..86R}{3}{Figure~2} of
\citeads{2012A&A...540A..86R}).}} 
underestimates their intensities and overestimates their limb
darkening.} 
Above the FALC temperature minimum the $B$ increase is not followed by
the flattening $J$ curves.
\reva{The $S$ curves sense $B$ somewhat but only} above
$\tau_\lambda\tis\,1$.

\revd{This photospheric control occurs similarly for ultraviolet
continua computed with the 1D FALP model, 
with network and plage brightening in the ultraviolet from the higher
deep-photosphere temperatures. 
Model-imposed ultraviolet MC brightening also holds for its
\textsf{SATIRE} modification which has no chromosphere and uses
$S\tis\,B$ contrast rather than NLTE-derived $S\tapprox\,J$ contrast
to mimic actual multi-D deep-hole radiation.}

The scattering $S-B$ splits at right translate with
Eq.~\ref{eq:WienS} into $b_c/b_l$ ratios shown at left as
logarithmic divergences. 
The metals are mostly ionized with most of the element in the ion
ground state so that $b_c\tapprox\,1$ (dashed) whereas the minority
$b_1$ curves \revc{(solid)} have substantial dips from 
\revcpar radiative over-ionization
and steep rises higher up from \revcpar radiative under\revc{-}ionization.
The corollary is that all lines of these species, in particular all
\FeI\ lines, have significant opacity depletion with respect to
\acp{LTE} which starts already in the deep photosphere.  
For the strongest lines it reverses into large over-opacity in the
FALC chromosphere.

The ionization of hydrogen is mostly from level $n\tis\,2$ so that the
Balmer continuum pattern at right \reva{defines} FALC hydrogen
ionization departures.  
Hydrogen is virtually neutral below the FALC transition region so that
$b_1\tapprox\,1$, and also \Lyalpha\ is virtually in detailed balance
so that $b_2\tapprox\,b_1\tapprox\,1$ \reva{(solid), 
giving \acp{LTE} extinction to the Balmer lines and continuum}.
Therefore $b_c$ (dashed) shows the Balmer over- and under-ionization
pattern reversely to $b_1$ of the metals.

\reva{A warning:} Figure~\ref{fig:edges} demonstrates scattering from radial
temperature gradients. 
In the actual time-dependent 3D solar atmosphere comparable ultraviolet $S-B$
scattering divergences occur for fine-structure gradients such as the
steep outward and lateral ones around granules, across hot walls in
\acp{MC}s, and around small-scale heating events. 
Any treatment short of detailed 3D$(t)$ radiative transfer is an
approximation. 

\begin{figure}      
  \centering
  \includegraphics[height=56mm]{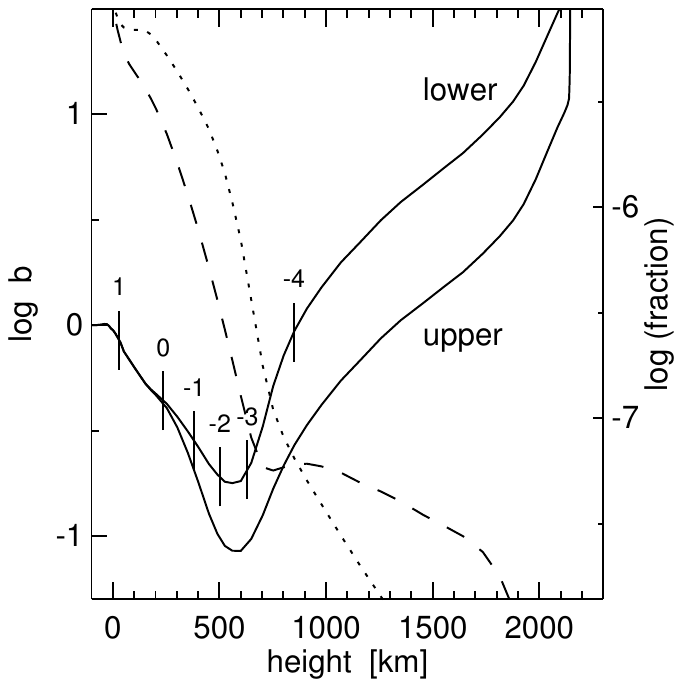}
  \includegraphics[height=56mm]{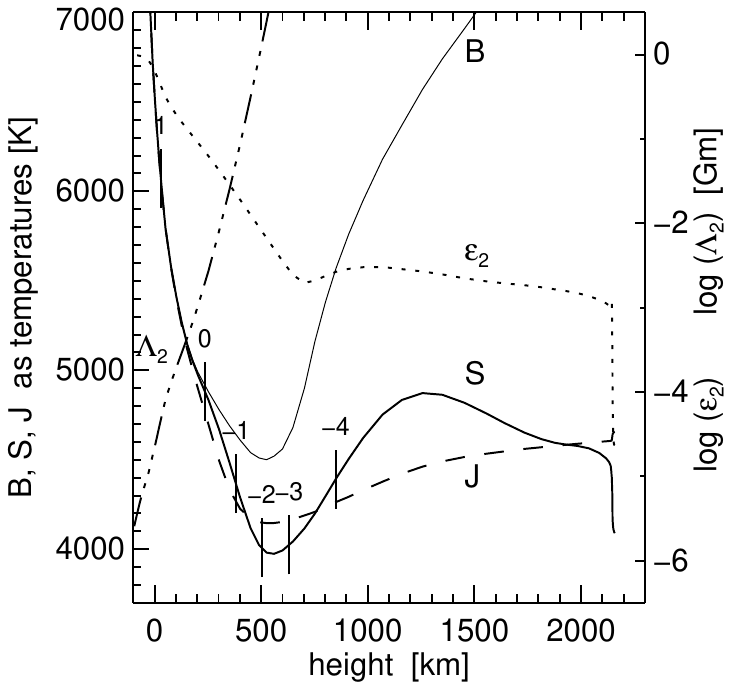}
  \caption[]{
  \FeI\ 6302\,\AA\ in the FALC star.
  {\em Left}: $b_u$ and $b_l$ ({\em solid\/}). 
  {\em Dashed\/}: population fraction $n_l/N_{\rm elem}$ (axis at
  right). 
  {\em Dotted\/}: same in \acp{LTE} (Saha--Boltzmann fraction). 
  {\em Right}: corresponding $B$ ({\em thin solid\/}), $J$ ({\em
  dashed\/}), and $S$ ({\em thick solid\/}) as formal temperatures. 
  {\em Dotted\/}: two-level collisional destruction probability
  $\varepsilon_2 = C_{ul}/(C_{ul}+A_{ul}+B_{ul}B)$ (scale at right). 
  {\em Dot-dashed\/}: two-level thermalization length
  $\Lambda_2 = \sqrt{\pi}/(\alpha_{\lambda_0} \varepsilon_2)$ for the
  Doppler core in gigameter\reva{s} (scale at right). 
  Example: $\reva{\log} \Lambda_2\tis\,-6$ implies thermalization of
  $S$ to $B$ at the center of a two-km thick feature.
  The curve label is placed besides the curve at the line-core
  thermalization height.
  The numbered ticks in both panels specify $\log \tau$
  heights for line center.  
  \reva{After}
  \href{http://www.staff.science.uu.nl/~rutte101/rrweb/rjr-edu/lectures/rutten_ssx_lec.pdf}{SSX}
  which \reva{offers} such graphs for many lines and multiple standard
  models.
  }
\label{fig:Fe6302}
\end{figure}

\begin{figure}      
  \centering
  \includegraphics[height=56mm]{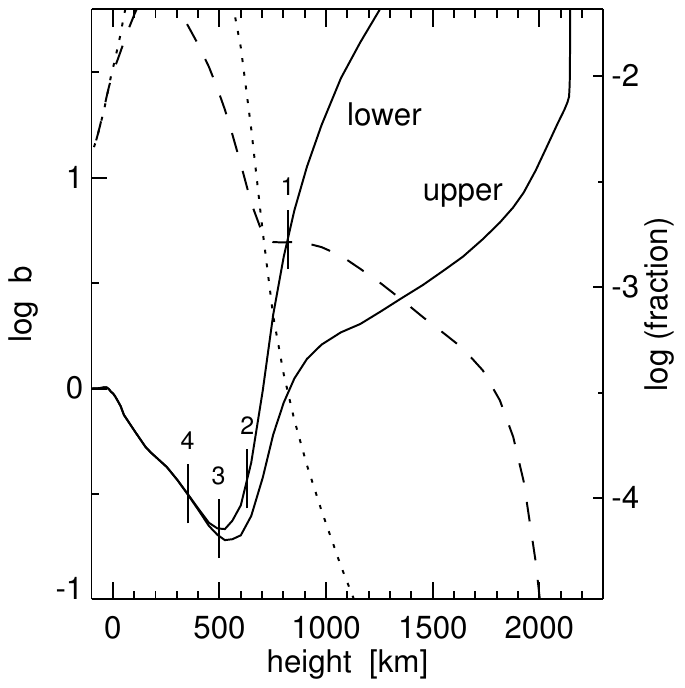}
  \includegraphics[height=56mm]{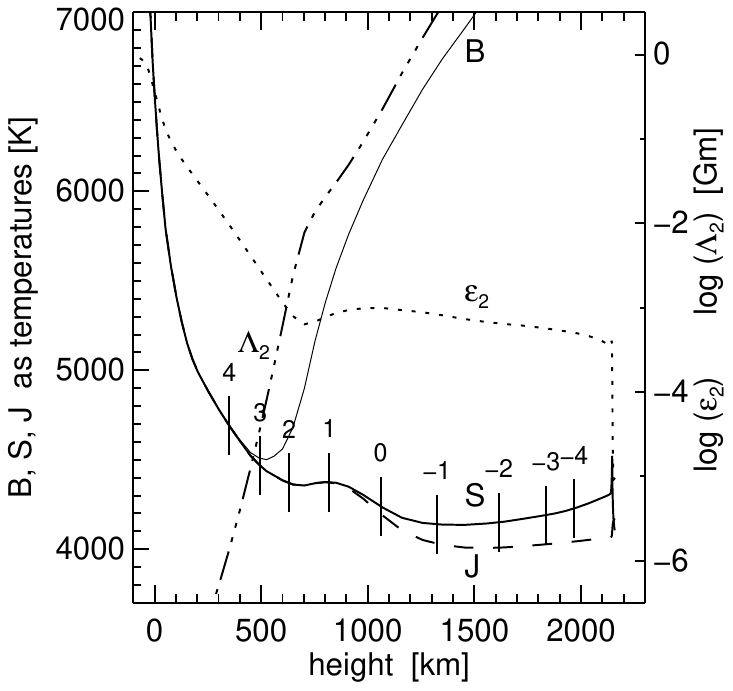}
  \caption[]{
  \FeI\ 3860\,\AA\ in the FALC star. 
  Format as Figure~\ref{fig:Fe6302}.
  }
\label{fig:Fe3860}
\end{figure}

\subsection{FALC Iron Lines}   \label{sec:lines}

Since irradiance studies are more concerned with the multitude of
spectral lines than with the specific lines employed in resolved solar
physics, the champion line producer is sampled here by showing FALC
formation for two \FeI\ lines, a weak optical one and a strong
near-ultraviolet one.

Figure~\ref{fig:Fe6302} is for the well-known optical polarimetry line
at 6302\,\AA.
The curve for two-level
$\varepsilon_2 = C_{ul}/(C_{ul}+A_{ul}+B_{ul}B)$ \reva{in the
right-hand panel} follows the electron density in
Figure~\ref{fig:FALC} \reva{and drops to $10^{-2}$} near
$\tau\tis\,1$, suggesting \reva{domination by} scattering \reva{in
Eq.~\ref{eq:S_CS}}. 
Thermalization occurs near $\tau\tis\,10$ in the deep photosphere, but
$S$ departs from $B$ only above $\tau\tis\,1$ to follow $J$ more
closely.
The reason for this apparent discrepancy is multi-level interlocking
(the $\eta$ terms in Eq.~\ref{eq:S_CS}) from the richness of the
\FeI\ Grotrian diagram.
This line is a fairly high-lying (multiplet 816, 3.6\,--\,5.6\,eV)
subordinate one among many others, as are most optical \FeI\ lines. 
Their levels are connected to lower levels by much stronger lines,
mostly in the ultraviolet, that de-thermalize ($S\tapprox\,J$
uncoupling from $B$) further out and so force $b_u\tis\,b_l$ equality
for their weaker siblings. 
Likewise, many weaker \FeI\ lines are members of multiplets in which
the strongest members impose similar thermalization.
The upshot is that most weak \FeI\ lines have source function
\acp{LTE} (``what'') thanks to the rich term structure.

However, they do not have extinction \acp{LTE} (``where''). 
This is evident from the $b_l$ curve \reva{in the left-hand panel}
which follows the ultraviolet-imposed pattern of photospheric dip and
chromospheric rise in Figure~\ref{fig:edges}. 
It translates into the divergence between the \acp{LTE} and
\reva{NLTE} 
fraction curves, whereas the $\reva{b_l\!-\!b_u}$ split corresponds to
the \reva{NLTE} $B\!-\!S$ split at right
(Eqs.~\ref{eq:Wienalpha} and \ref{eq:WienS}).
\revapar Polarimetry ``inversions'' \reva{with this line and/or
similar ones} often obtain NLTE source function estimates from
best-fit modeling \reva{but ignore} the more important $b_l$ dip in
the extinction.

Figure~\ref{fig:Fe3860} is in the same format but for a strong
near-ultraviolet \FeI\ line, member of multiplet 4 from the ground
state.
The latter has fractional population (dashed curve at left) about
$10^{-2}$ or less, confirming that iron is predominantly ionized
everywhere. 
The Saha--Boltzmann value (dotted) is significantly higher in the
photosphere and lower in the FALC chromosphere because the $b_l$ curve
is the $b_1$ curve of Figure~\ref{fig:edges}
\revapar 
set by ultraviolet bound--free scattering.
The $b_u$ curve drops away from it following the $B\!-\!S$ split at
right. 
This line scatters more strongly ($\varepsilon_2$ below $10^{-3}$) and
behaves more as a two-level line, with thermalization near
$\tau\tis\,10^3$ and $S\tapprox\,J$ separation from $B$ starting
already there, far below its $\tau\tis\,1$ \reva{height}, and with
only \reva{marginal} 
sensitivity to the FALC temperature rise so that the emergent profile
is an absorption dip without core reversal.
One might call the line ``chromospheric'' because it \reva{has}
$\tau\tis\,1$ above $h\tis\,1000$\,km, but its core intensity responds
rather to temperature modulation of the FALC temperature minimum where
the emerging photons are created. 
However, line-core Doppler and Zeeman measures are imposed at the last
scattering and do respond higher up.

\begin{figure}      
  \centering
  \includegraphics[width=\textwidth]{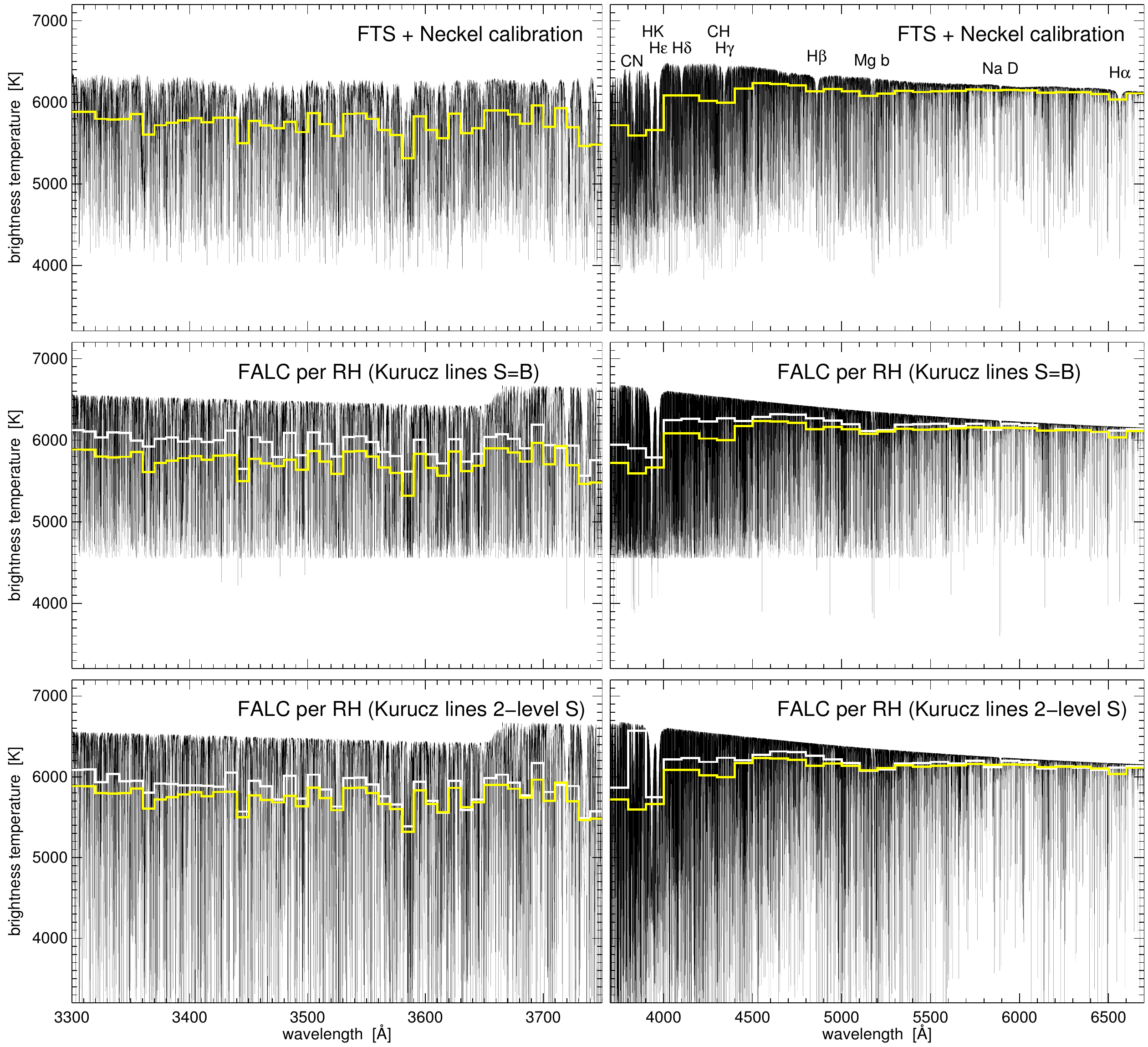}
  \caption[]{
  Line haze in the near-ultraviolet ({\em left}) and optical ({\em
  right}, with major features identified along the top).  
  The \reva{left and right wavelength scales differ in dispersion and
  overlap over 3700\,--\,3750\,\AA.}
  The \reva{overlaid} histograms are averages over 10\,\AA\ bins at
  left, 100\,\AA\ bins at right. 
  \reva{The yellow histograms hold for the observations in the
  top panels and are replicated in the lower panels to facilitate
  comparison.}
  {\em Top row}: disk center intensity spectra taken with the Kitt
  Peak Fourier Transform Spectrometer by J.W.~Brault, calibrated into
  absolute intensity by \citetads{1984SoPh...90..205N} 
  and posted by
  \citetads{1999SoPh..184..421N}, 
  converted to brightness temperature to remove the Planck-function
  sensitivity pattern.
  {\em Middle row}: \reva{spectra} synthesized with \textsf{RH} using
  \acp{LTE} \reva{at} all wavelengths sampling Kurucz lines instead of
  explicit lines.
  {\em Bottom row}: \reva{spectra} synthesized with \textsf{RH} using
  monochromatic two-level scattering \reva{at all} wavelengths
  sampling atomic Kurucz lines.
  \reva{The high 3800\,--\,3900 histogram value is an artifact.}  
  }
  \label{fig:haze}
\end{figure}

\section{Obstacles in Simulation-based Irradiance Modeling}
\label{sec:obstacles}

\subsection{(Ultra)violet Line Haze} \label{sec:linehaze}

The multitude of weak and strong lines
\reva{displayed in the top curve of Figure~\ref{fig:extinction} and}
exemplified by
Figures~\ref{fig:Fe6302} and \ref{fig:Fe3860} together constitute a
dense ``line haze'' in the blue, violet, and ultraviolet
(\citeads{1972SoPh...22...64L}, 
\citeads{1980A&A....90..239G}) 
that must be included in spectral synthesis for irradiance modeling.
Figures~\ref{fig:Fe6302} and \ref{fig:Fe3860} demonstrate that their
scattering nature requires \acp{NLTE} treatment.

In principle, the proper way is the brute force method: 
include all pertinent levels
and lines in comprehensive model atoms that constitute detailed input
for a \acp{NLTE} spectral synthesis code handling each line
explicitly.
In practice, this is feasible for single-pixel plane-parallel 1D
modeling as shown by
\citetads{2005ApJ...618..926S} 
and \citetads{2015ApJ...809..157F}, 
but it remains undoable for 3D time-dependent simulations with 3D
time-dependent radiative transfer -- defining a need for tractability
simplifications. 

\reva{Statistical methods using grouping of levels and lines of dominating
elements and stages were pioneered by
\citetads{1989ApJ...339..558A} 
and used in constructing line-blanketed stellar atmosphere models (see
\linkadspage{2003ASPC..288...31W}{18}{Section~6} in the review by
\citeads{2003ASPC..288...31W}). 

Yet simpler recipes are used in spectrum synthesis codes
commonly used in solar modeling.}
\revapar \reva{Avrett's} \textsf{Pandora} \reva{code applies a simple recipe}
\revapar detailed on
\linkadspage{2008ApJS..175..229A}{15}{p.~243\,ff.} of
\citetads{2008ApJS..175..229A}. 
It forces a gradual transition from $S\tis\,B$ in the deep photosphere
to $S\tis\,J$ in the model chromosphere for all lines in the
\citetads{2009AIPC.1171...43K} 
list that are not explicitly solved. 
This imposed \reva{$B \rightarrow J$} transition is the same for all Kurucz
lines and is derived from trial fits for each model atmosphere.

\reva{The \textsf{RH} code} currently offers three options. 
The first is the Zwaan-inspired Bruls recipe of wavelength-dependent
increase of the metal and \Hmin\ extinctions with the best-fit
multipliers shown in \linkadspage{1992A&A...265..237B}{3}{Figure~2} of
\citetads{1992A&A...265..237B}. 
These tables were derived with quiet-Sun data and a
quiet-Sun model and are therefore applicable only in such modeling. 

The second \reva{\textsf{RH}} option is detailed sampling of the line
list of \citetads{2009AIPC.1171...43K} 
with \acp{LTE} source-function evaluation.
The third \reva{\textsf{RH} option} is \reva{to use} \revapar
\reva{line source function evaluation assuming} \revapar monochromatic
two-level scattering \revapar using Eq.~\ref{eq:S_CS} \reva{with
$\eta\tis\,0$ and} \revapar the Van Regemorter estimate
(\linkrtsa{70}{Eq.~3.32}) for $\varepsilon$ \reva{in
\lambdop-iteration at} all \reva{wavelengths sampling} atomic lines
\reva{in the Kurucz list} while maintaining $S_\lambda\tis\,B_\lambda$
for the molecular lines.  
\reva{The second and third options are also offered for specified
background opacities in the \textsf{MULTI} code of
\citetads{1986UppOR..33.....C} 
and the \textsf{Multi3D} variant of
\citetads{2009ASPC..415...87L}.} 

The Avrett and \reva{third} \textsf{RH} recipes \reva{address
scattering} in \revapar line-haze source functions but the extinctions
are still evaluated per Saha--Boltzmann, for minority species ignoring
the under- and over-ionization pattern in Figure~\ref{fig:edges}.
These recipes also ignore over-excitation through ultraviolet
multi-level pumping as \reva{often occurs} in \FeII\
(\citeads{1988ASSL..138..185R}), 
also a major line-haze contributor.

Figure~\ref{fig:haze} demonstrates \reva{the second and third}
\textsf{RH} \revapar options.
\revb{The top row shows that the observed disk-center brightness
temperature peaks around 6500\,K at 4000\,\AA\ where one probes the
solar atmosphere effectively as deep as in the $1.6\,\mu$m opacity
minimum (\citeads{1989SoPh..124...15A}).} 
\revcpar

In the second row, strong Kurucz lines cannot reach deeper than the
FALC 4400\,K minimum temperature and obtain core reversals.
The lines that do reach lower were specified in the model atoms and
treated explicitly. 
These include the labeled atomic lines in the top-right panel.
The \NaID\ (darkest), \MgIb, \CaIIHK\ (in \acp{PRD}), and Fe lines
(including \revapar \reva{6302}\,\AA\ and 3860\,\AA) are well
reproduced, \reva{but} the Balmer lines have insufficiently extended
wings (upper-envelope dips in the top panel) because \textsf{RH} does
not apply the Holtsmark distribution for linear Stark broadening, also
resulting in lack of merged line blanketing towards the Balmer limit
\reva{at 3646\,\AA}.

\reva{In both wavelength ranges,} many more Kurucz lines reach the
\reva{temperature-minimum} threshold than in the observed spectra in
the first row, likely due to the neglect of radiative over-ionization. 
\reva{They are all too strong, but nevertheless the mean histograms
lie increasingly above the observed ones towards shorter wavelength,
suggesting yet more lines than in the Kurucz list (assuming that the
observation calibration is correct).}
\revb{This is also suggested by the larger raggedness of the upper
envelope for the observations.}

The bottom row shows \textsf{RH}'s monochromatic-scattering result.
\reva{The lower envelope at right resembles the observed one towards
longer wavelength, with many computed lines reaching lower brightness
temperatures than the minimum temperature through scattering.
However, many computed lines reach deeper than observed, more in the
blue and yet more so in the ultraviolet at left.}
This is not only due to ignoring radiative over-ionization in the line
extinctions but also because the monochromatic two-level approximation
ignores the interlocking which brings weaker members of multiplet and
term \reva{groups} closer to $S\tis\,B$ as in Figure~\ref{fig:Fe6302}. 
\reva{The mean histograms are closer to the yellow observed ones than
in the middle row, but they still lie above these.  
Even with too much scattering the line haze remains underestimated.}

Clearly, \reva{these} recipes are unsatisfactory. 
I suggest \textsf{RH} experiments with the following tractability
simplification: construct a relatively small but representative model
atom for element ``fudge'' with Fu\,{\sc i} and Fu\,{\sc ii}
resembling \FeI\ and \FeII\ \reva{including strong ultraviolet and
weaker optical subordinate lines}, solve its transitions explicitly,
and \reva{then} apply the resulting population departures $b$ sorted
per excitation energy to both $\alpha^l$ and $S^l$ of all atomic
Kurucz lines.
\reva{This scheme represents a step up from Avrett's all-the-same
recipe but is simpler than grouping into superlevels per species.
It may give better reproduction of the Kurucz lines than the two-level
scattering in Figure~\ref{fig:haze} or when using Avrett's recipe, but
it will not solve the apparent incompleteness of the Kurucz list}.

\begin{figure}[t]
  \centerline{\includegraphics[width=\textwidth]{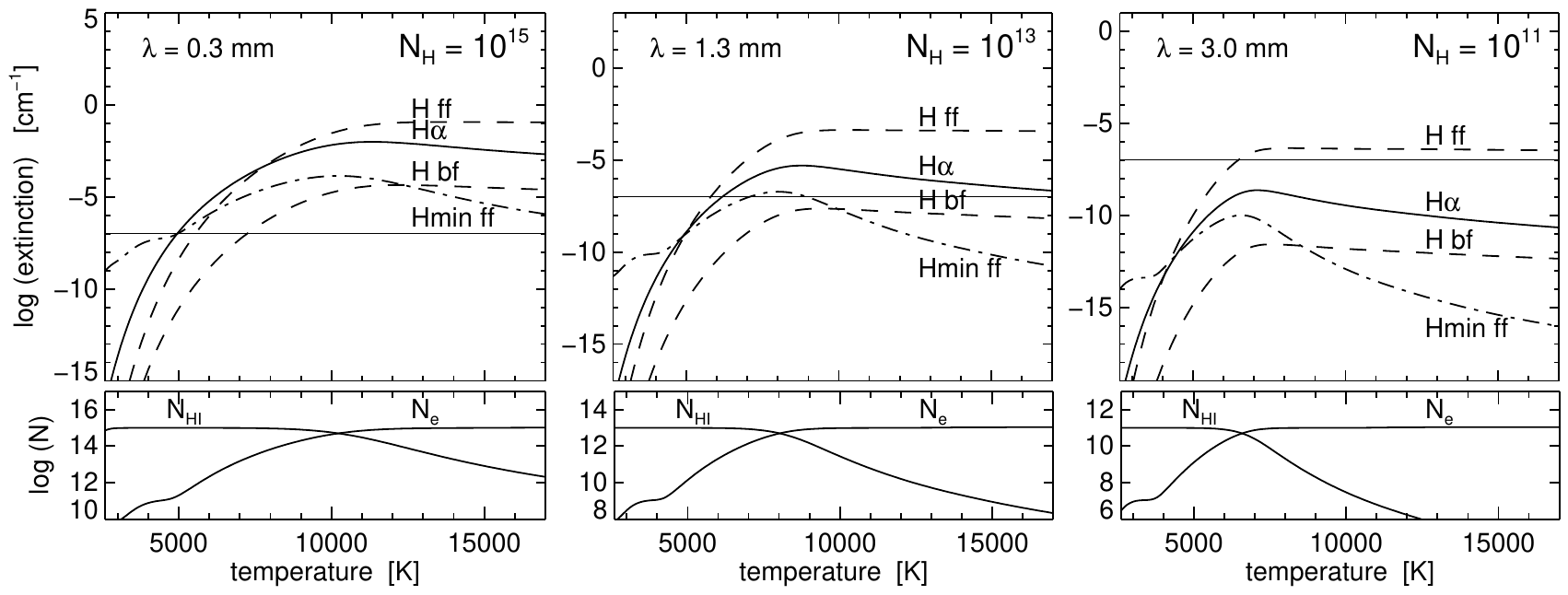}}
  \caption[]{
  {\em Upper panels}: Saha--Boltzmann line extinction coefficient as
  function of temperature at the center of \Halpha\ ({\em solid\/})
  and continuous extinction coefficient of the \HI\ free--free and
  bound--free contributions ({\em dashed\/}) and the \Hmin\ free--free
  contribution ({\em dot-dashed}) at three \acp{ALMA} wavelengths
  (from left to right 0.35, 1.3, and 3.0\,mm), for gas of solar
  composition with total hydrogen densities
  $N_\rmH\tis\,10^{15}, 10^{13}$, and $10^{11}$~cm$^{-3}$.
  The latter correspond to FALC radiation escape height for these
  wavelengths at the bottom, middle, and top of the FALC chromosphere
  (Figure~\ref{fig:FALC}).
  The horizontal line at $y\tis\,-7$ shows the extinction at which a
  100\,km thick feature becomes optically thick.
  {\em Lower panels}: competing neutral hydrogen and electron
  densities (cm$^{-3}$), with the same logarithmic unit size as the
  extinction scales to enable slope comparisons.   
  The $y$ scales including the $y\tis\,-7$ line shift up per column
  whereas the extinction maxima shift left and down.
  After \linkadspage{2017A&A...598A..89R}{4}{Figure~1} of
  \citetads{2017A&A...598A..89R}. 
  }
  \label{fig:extsb}
\end{figure}

\subsection{Non-E  mm Radiation} \label{sec:nonEmm}

At wavelengths above 0.1\,mm non-E time dependence is likely important
in solar continuum formation and spectral irradiance because this
radiation samples domains where hydrogen ionizes in static 1D modeling
(Figure~\ref{fig:FALC}).
Shorter infrared wavelengths may already sample dynamic hydrogen
ionization \reva{occurring} in lower-lying small-scale heating events
\reva{such as the ``rapid blue and red excursion'' features in \Halpha\
that are on-disk manifestations of Type-II spicules}
(\citeads{2009ApJ...705..272R}; 
 \citeads{2013ApJ...769...44S}). 

The mm region was so far underobserved but the advent of solar
observing with the {\em \revc{\acl{ALMA}}} (ALMA) promises ``game-changing''
results, in
particular when \acp{ALMA} development permits long-baseline modes
potentially yielding higher angular resolution even than the
SST. \revapar

At these wavelengths the ``what'' source function question is easy
because the extinction is primarily free--free (\Hmin\,ff in the
upper photosphere, transiting to \HI\,ff wherever hydrogen ionizes) so
that $S\tis\,B$ and since the Rayleigh-Jeans approximation also holds
$I_\lambda\tapprox\,T_\rme(\tau_{\lambda\mu}\tis\,1)$ for optically
thick features while thin features give
$\Delta I_\lambda\tapprox\,\tau_{\lambda\mu}T_\rme$ contributions.
With proper calibration \acp{ALMA} directly delivers temperatures.

However, the ``where'' question concerning $\tau_{\lambda\mu}$ is
intricate because both hydrogen free--free extinctions depend on
hydrogen ionization: \Hmin\,ff vanishes at it, \HI\,ff requires it. 
This is shown by comparing the \reva{\Hmin\,ff} and \reva{\HI\,ff}
curves in the upper panels of Figure~\ref{fig:extsb} with the
ionization curves in the lower panels.
In all three columns \Hmin\,ff dominates only below 5000\,K, with a
plateau from electron-donor ionization also present in the
$N_\rme$ curves.
\HI\,ff extinction increases Boltzmann-steep with temperature and
saturates for
full hydrogen ionization ($N_\rme\tapprox\,N_{\rmH}$) at very high values.
Its offset above \Halpha\ extinction increases $\sim \lambda^2$
(\linkrtsa{47}{Eq.~2.79}). 

The extinction coefficients in Figure~\ref{fig:extsb} are computed
assuming Saha--Boltz\-mann partitioning.
For all, including \Halpha, this is a good assumption at high
temperature for which collisions up and collisions down in the
\Lyalpha\ transition balance fast.
However, in dynamic situations where heated gas cools drastically this
balancing becomes very slow because the 10\,eV \Lyalpha\ jump is so
large.
The retarded $n\tis\,2$ population then stays high \revc{initially},
translating into \revapar \Halpha\ over-extinction.
In terms of the \Halpha\ curves in Figure~\ref{fig:extsb}:
\reva{during heating events} the \Halpha\ extinction \reva{obeys
Saha--Boltzmann and} rises steeply along the curves towards their tops
at 80\,\% hydrogen ionization, but in cooling aftermaths it does not
instantaneously follow the temperature back down along the
Saha--Boltzmann curves but hangs during minutes of retardation near
its high previous values.

\reva{The $n\tis\,2$} retardation \revapar affects hydrogen ionization
since that is governed by the $n\tis\,2$ population in
Balmer-continuum loops \reva{as in Figure~\ref{fig:detours}},
\revapar
of which the domination was demonstrated in
\linkadspage{2002ApJ...572..626C}{4}{Figure~3} of
\citetads{2002ApJ...572..626C}. 
This ionization loop operates in instantaneous \acp{SE} and modulates
$b_c$ over 1\,--\,2 orders of magnitude as in Figure~\ref{fig:edges},
but this modulation represents only a minor addition to the 5\,--\,10
orders of magnitude change along the steep Boltzmann slope \reva{in
Figure~\ref{fig:extsb}} that defines \Lyalpha-settling retardation.
Non-E retardation excesses of this very large size have been well
documented for acoustic inter-network shocks by
\citetads{2002ApJ...572..626C} 
\reva{and
\citetads{2007A&A...473..625L},} 
and also for \acp{MC}-guided shocks producing dynamic \Halpha\ fibrils
above network by
\citetads{2007A&A...473..625L}. 
Such gigantic \reva{$n\tis\,2$} over-populations in dynamically
cooling gas \revd{probably}
contribute to the excessively rich fibrilar scenes
observed in \Halpha, not only with shocks as a prior heating agent but
also \reva{with} \revapar small-scale reconnection \reva{events}
(\citeads{2017ApJ...847...36M}). 
Reconnection probably also caused the exemplary ``contrail'' fibril of
\citetads{2017A&A...597A.138R}. 

\reva{\Halpha\ features that gain non-E opacity in rapid cooling after
dynamic heating will similarly have retarded non-E opacity} at
\acp{ALMA} wavelengths since the H\,ff curves in
Figure~\ref{fig:extsb} share the very steep $n\tis\,2$ Boltzmann rise
of the \Halpha\ curves and then level out above \Halpha.
\revb{The upshot is that for ALMA continua the source function is
strictly local in
space-time, but as for \Halpha\ the extinction is non-local in time
from retarded \Lyalpha\ settling and in space from \Lyalpha\
irradiation.}
High-resolution \acp{ALMA} images \revb{may} not look like \Halpha\
images because the source functions (``what'') are set
\reva{discordantly} by $B_\lambda$ and $J_\lambda$, respectively, but
non-E \Halpha\ features will show up with \reva{even} larger optical
thickness (``where'') \reva{with ALMA}.
Figure~\ref{fig:extsb} shows that at the top of the FALC chromosphere
(right-most column) a hot -- or a post-hot -- 100-km thin feature becomes
fully transparent in \Halpha\ but remains optically thick and
measurable with \acp{ALMA} at 3\,mm, hopefully \reva{eventually} also
resolvable. 
For more detail see Rutten
(\citeyearads{2017IAUS..327....1R}, 
\citeyearads{2017A&A...598A..89R}). 

\reva{Standard 1D models cannot replicate \Halpha\ images and are
therefore} useless to interpret \acp{ALMA} imaging even for
irradiance interests.
\acp{ALMA} interpretation requires non-E numerical simulations that
furnish realistic \Halpha\ scenes to begin with.
However, non-E 3D time-dependent simulation including non-E 3D
time-dependent spectral synthesis\footnote{The need for 3D in \Halpha\
scenes was beautifully demonstrated in
\linkadspage{2012ApJ...749..136L}{7}{Figure\,7} of
\citetads{2012ApJ...749..136L}. 
Their \acp{MHD} simulation was non-E
(\citeads{2016A&A...585A...4C}), 
but their \Halpha\ synthesis still assumed instantaneous \acp{SE}.}
remains undoable at present\footnote{\reva{SE 3D time-dependent
synthesis is already challenging, see
\citetads{2019AdSpR..63.1434P}.}}. 

I suggest the following tractability simplification: do not compute
non-E-retarded hydrogen $n\tis\,2$ populations in full detail but
\reva{store} for each parcel of gas the highest \reva{value} of
Saha--Boltzmann $n_2/N_\rmH$ that it \reva{reached} in
the past minutes \reva{and maintain that ratio} during cool aftermaths,
or use the peak values with retarded decay along the extinction curves
in Figure~\ref{fig:extsb}. 
This recipe requires parcel labeling for tracing its whereabouts as in
\citetads{2018A&A...616A.136L}. 

\section{Conclusion} \label{sec:conclusion}
\reva{Solar irradiance modeling of the contributions by network and
plage is presently in an important and timely transition from questionable
classic 1D fitting to more realistic simulation-based interpretation.
The ultraviolet line haze requires detailed NLTE evaluation because it
controls the NLTE opacity departures of most atomic lines throughout
the spectrum.  
At long wavelengths non-E hydrogen ionization and recombination
produce large continuum opacities in gas that cools after dynamical
heating.
Both complexities are challenging; I suggest experiments with the
tractability recipes given above.
}

\begin{acks}
I thank the organizers of Focus Meeting FM9 at the 30th General
Assembly of the \acp{IAU} for inviting me to review this topic and so
triggering this publication.
\reva{I thank the reviewer for suggesting many improvements and ADS
for assistance with its page serving.}
\end{acks}

\end{article} 
\end{document}

%% file: rrmacros-pub.tex
\definecolor{bllsht}{rgb}{0.40, 0.05, 0.01} 
\long\def\@ #1@{\par\vspace{0.5ex}\noindent
  {\Large \textcolor{red}{@~}}\color{bllsht}{#1}\color{black}\\[0.5ex]}
 
\makeatletter
\newcommand{\bibnote}[2]{\global\@namedef{#1note}{#2~}}
\newcommand{\biblink}[2]{\global\@namedef{#1link}{#2}}
\makeatother

\long\def\reva#1{{\color{red}{\bf #1}\color{black}}}  
\def\revapar{{\color{red}{\P\ }\color{black}}} 
\definecolor{amber}{rgb}{1.0, 0.49, 0.0}
\long\def\revb#1{{\color{amber}{\bf #1}\color{black}}}  
\def\revbpar{{\color{amber}{\P\ }\color{black}}} 
\long\def\revc#1{{\color{red}{\bf #1}\color{black}}}  
\def\revcpar{{\color{red}{\P\ }\color{black}}} 
\long\def\revd#1{{\color{amber}{\bf #1}\color{black}}}  

\bibpunct{(}{)}{;}{a}{}{,}    
\makeatletter
  \protected\def\sppresslink{\def\hyper@linkstart##1##2{}\let\hyper@linkend\@empty}
  \newcommandtwoopt{\citeads}[3][][]{%
   \href{http://ui.adsabs.harvard.edu/abs/#3/abstract}%
        {\sppresslink \citealp[#1][#2]{#3}}
   \biblink{#3}{\href{http://ui.adsabs.harvard.edu/abs/#3/abstract}{ADS}}}
 \newcommandtwoopt{\citepads}[3][][]{%
   \href{http://ui.adsabs.harvard.edu/abs/#3/abstract}%
        {\sppresslink \citep[#1][#2]{#3}}
   \biblink{#3}{\href{http://ui.adsabs.harvard.edu/abs/#3/abstract}{ADS}}}
 \newcommandtwoopt{\citetads}[3][][]{%
   \href{http://ui.adsabs.harvard.edu/abs/#3/abstract}%
        {\sppresslink \citet[#1][#2]{#3}}
  \biblink{#3}{\href{http://ui.adsabs.harvard.edu/abs/#3/abstract}{ADS}}}
 \newcommandtwoopt{\citeyearads}[3][][]{%
   \href{http://ui.adsabs.harvard.edu/abs/#3/abstract}%
        {\sppresslink \citeyear[#1][#2]{#3}}
   \biblink{#3}{\href{http://ui.adsabs.harvard.edu/abs/#3/abstract}{ADS}}}
\makeatother

\def\linkadspage#1#2#3{\href{http://adsabs.harvard.edu/cgi-bin/nph-data_query?bibcode=#1\&link_type=ARTICLE\&db_key=AST\#page=#2}{#3}}

\newcount\longrefs
\def\adv{\ifnum\longrefs=1 {Adv.\ Space Res.} \else 
                           {Adv.\ Sp'\ Res.}\fi}
\def\aap{\ifnum\longrefs=1 {Astron.\ Astrophys.}\else 
                           {A\hbox{\rm \&}A}\fi}
\def\aapr{\ifnum\longrefs=1 {Astron.\ Astrophys.\ Rev.}\else 
                            {A\hbox{\rm \&}AR}\fi}
\def\aaps{\ifnum\longrefs=1 {Astron.\ Astrophys.\ Suppl.}\else 
                            {A\hbox{\rm \&}A Suppl.}\fi}
\def\actaa{\ifnum\longrefs=1 {Acta Astronomica}\else
                            {Acta Astron.}\fi}
\def\aipcs{\ifnum\longrefs=1 {Am.\ Inst.\ Phys.\ Conf.\ Series}\else
                             {AIP Conf.\ Ser.}\fi}
\def\aj{\ifnum\longrefs=1 {Astron.\ J.}\else 
                          {AJ}\fi} 
\def\ao{\ifnum\longrefs=1 {Applied Optics}\else 
                           {Appl.\ Opt.}\fi} 
\def\aspcs{\ifnum\longrefs=1 {Astron.\ Soc.\ Pacific Conf.\ Series}\else 
                           {ASP Conf.\ Ser.}\fi} 
\def\apj{\ifnum\longrefs=1 {Astrophys.\ J.}\else 
                           {ApJ}\fi} 
\def\apjl{\ifnum\longrefs=1 {Astrophys.\ J. Lett.}\else 
                            {ApJL}\fi} 
\def\aplett{\ifnum\longrefs=1 {Astrophys.\ J. Lett.}\else 
                            {ApJ}\fi} 
\def\apjs{\ifnum\longrefs=1 {Astrophys.\ J. Suppl.}\else 
                            {ApJS}\fi}
\def\apss{\ifnum\longrefs=1 {Astrophys.\ Space Sci.}\else 
                            {Astrophys.\ Space Sci.}\fi}
\def\araa{\ifnum\longrefs=1 {Ann.\ Rev.\ Astron.\ Astrophys.}\else 
                            {ARA\hbox{\rm \&}A}\fi}
\def\azh{\ifnum\longrefs=1 {Astronomicheskii Zhurnal}\else 
                            {Astron.\ Zhur.}\fi}
\def\baas{\ifnum\longrefs=1 {Bull.\ Am.\ Astron.\ Soc.}\else 
                            {BAAS}\fi}
\def\bain{\ifnum\longrefs=1 {Bull.\ Astronom.\ Institutes Netherlands}\else
                            {Bull.\ Astr.\ Inst.\ Neth.}\fi}
\def\cjaa{\ifnum\longrefs=1 {Chin.\ J.\ Astron.\ Astrophys.}\else 
                            {Chin.\ J.\ A\&A}\fi}
\def\gca{\ifnum\longrefs=1 {Geochim.\ Cosmochim.\ Acta}\else 
                           {Geochim.\ Cosmochim.\ Acta}\fi}
\def\grl{\ifnum\longrefs=1 {Geophys.\ Res.\ Lett.}\else 
                           {Geoph.\ Res.\ Lett.}\fi}
\def\iaucirc{\ifnum\longrefs=1 {IAU Circulars}\else 
                          {IAU Circ.}\fi}
\def\icarus{\ifnum\longrefs=1 {Icarus}\else 
                          {Icarus}\fi}
\def\ip{\ifnum\longrefs=1 {in press}\else 
                          {in press}\fi}
\def\jcap{\ifnum\longrefs=1 {Jour.\ Cosmology Astropart.\ Phys.}\else 
                          {JCAP}\fi}
\def\jgr{\ifnum\longrefs=1 {J.\ Geophys.\ Res.}\else 
                           {J.\ Geophys.\ Res.}\fi}  
\def\jrasc{\ifnum\longrefs=1 {J.\ Royal Astron.\ Soc.\ Canada}\else 
                             {JRAS Can.}\fi}  
\def\memsai{\ifnum\longrefs=1 {Mem.~Soc.~Astron.~Italiana}\else
                              {MmSAI}\fi}
\def\mnras{\ifnum\longrefs=1 {Mon.\ Not.\ Roy.\ Astron.\ Soc.}\else 
                             {MNRAS}\fi} 
\def\na{\ifnum\longrefs=1 {New Astronomy}\else 
                          {New Astron.}\fi}
\def\nar{\ifnum\longrefs=1 {New Astronomy rev.}\else 
                           {New Astron.\ Rev.}\fi}
\def\nat{\ifnum\longrefs=1 {Nature}\else 
                           {Nat}\fi}
\def\pasa{\ifnum\longrefs=1 {Pub.\ Astron.\ Soc.\ Australia}\else 
                            {PASA}\fi} 
\def\pasj{\ifnum\longrefs=1 {Pub.\ Astron.\ Soc.\ Japan}\else 
                            {PASJ}\fi} 
\def\pasp{\ifnum\longrefs=1 {Pub.\ Astron.\ Soc.\ Pacific}\else 
                            {PASP}\fi} 
\def\physscr{\ifnum\longrefs=1 {Physica Scripta}\else 
                               {Phys.\ Scrip.}\fi} 
\def\planss{\ifnum\longrefs=1 {Planetary \& Space Science}\else 
                              {Plan. \& Space Sci.}\fi} 
\def\pre{\ifnum\longrefs=1 {Phys.\ Rev.\ E}\else
                           {Phys.\ Rev.\ E}\fi}
\def\procspie{\ifnum\longrefs=1 {Proc.\ SPIE}\else 
                                {Proc.\ SPIE}\fi} 
\def\qjras{\ifnum\longrefs=1 {Quarterly J.\ Royal Astron.\ Soc.}\else 
                             {QJRAS}\fi} 
\def\rmxaa{\ifnum\longrefs=1 {Revista Mexicana de Astron.\ y Astrofys.}\else 
                             {RMxAA}\fi} 
\def\sa{\ifnum\longrefs=1 {Soviet Astron..}\else 
                          {Sov.\ Astron.}\fi}
\def\skytel{\ifnum\longrefs=1 {Sky \& Telescope}\else 
                              {Sky \& Tel.}\fi} 
\def\solphys{\ifnum\longrefs=1 {Solar Phys.}\else 
                               {SoPh}\fi}
\def\sovast{\ifnum\longrefs=1 {Soviet Astron.}\else 
                              {Sov.\ Ast.}\fi}
\def\ssr{\ifnum\longrefs=1 {Space Sci.\ Rev.}\else 
                           {Space Sci.\ Rev.}\fi}
\def\zap{\ifnum\longrefs=1 {Zeitschr.\ f.\ Astrophysik}\else
                               {Z.\ Astrophys.}\fi}

\def\nl{,\ } 

\def\ITA{Institute of Theoretical Astrophysics\nl
         University of Oslo\nl
         P.O.\ Box 1029 Blindern\nl
         N-0315 Oslo\nl
         Norway}

\def\LA{Lingezicht Astrophysics\nl 't Oosteneind 9\nl 4158\,CA Deil\nl 
        The Netherlands}

\def\RoCS{Rosseland Centre for Solar Physics\nl
          University of Oslo\nl
          P.O.\ Box 1029 Blindern\nl
          N-0315 Oslo\nl
          Norway}

\def\acp#1{#1} 
\newacro{AA}{Astronomy \& Astrophysics}  
\newacro{ADS}{Astrophysics Data System}
\newacro{AIA}{Atmospheric Imaging Assembly}
\newacro{ALMA}{Atacama Large Millimeter/submillimeter Array}
\newacro{AO}{adaptive optics}
\newacro{ApJ}{Astrophysical Journal}
\newacro{AR}{active region}
\newacro{bb}{bound-bound}
\newacro{bf}{bound-free}
\newacro{BFI}{Broad-band Filter Imager}
\newacro{CE}{coronal equilibrium}
\newacro{CfA}{Center for Astrophysics}
\newacro{CME}{coronal mass ejection}
\newacro{CRD}{complete redistribution}
\newacro{CRISP}{CRisp Imaging SpectroPolarimeter}
\newacro{CRISPEX}{CRisp SPectral EXplorer}
\newacro{CS}{coherent scattering}
\newacro{DEM}{Differential Emission Measure}
\newacro{DF}{dynamic fibril}
\newacro{DKIST}{Daniel K. Inouye Solar Telescope}
\newacro{DLR}{Deutsches Zentrum f\"ur Luft- und Raumfahrt}
\newacro{DOT}{Dutch Open Telescope}
\newacro{DST}{Richard B. Dunn Solar Telescope}   
\newacro{EB}{Ellerman bomb}
\newacro{EDP}{\'{E}dition Diffusion Presse}  
\newacro{EIT}{Extreme ultraviolet Imaging Telescope}
\newacro{EPIC}{European participation in Solar-C}
\newacro{ERC}{European Research Council}
\newacro{ESA}{European Space Agency}
\newacro{EST}{European Solar Telescope}
\newacro{EUV}{extreme ultraviolet}
\newacro{FAF}{flaring active-region fibril}
\newacro{ff}{free-free}
\newacro{FITS}{Flexible Image Transport System}
\newacro{FOV}{field of view}
\newacro{fov}{field of view}
\newacro{FWHM}{full width at half maximum}
\newacro{HAO}{High Altitude Observatory}
\newacro{HD}{hydrodynamics}
\newacro{Hi-C}{High Resolution Coronal Imager Sounding Rocket}
\newacro{HMI}{Helioseismic and Magnetic Imager}
\newacro{IAA}{Instituto de Astrof\'{i}sica de Andaluc\'{i}a}
\newacro{IAC}{Instituto de Astrof\'{i}sica de Canarias}
\newacro{IAS}{Institut d'Astrophysique Spatiale}
\newacro{IAU}{International Astronomical Union}
\newacro{IBIS}{Interferometric Bi-dimensional Spectrometer}
\newacro{IDL}{Interactive Data Language}
\newacro{IMaX}{Imaging Magnetograph eXperiment}
\newacro{INAF}{Istituto Nazionale di Astrofisica}
\newacro{IB}{IRIS bomb}
\newacro{IR}{infrared}
\newacro{IRIS}{Interface Region Imaging Spectrograph}
\newacro{ISAS}{Institute of Space and Astronautical Science}
\newacro{ISP}{Institute for Solar Physics}
\newacro{ISS}{International Space Station}
\newacro{ISSI}{International Space Science Institute}
\newacro{ITA}{Institute for Theoretical Astrophysics}
\newacro{JAXA}{Japan Aerospace Exploration Agency}
\newacro{JSOC}{Joint Science Operations Center}
\newacro{KIS}{Kiepenheuer--Institut f\"{u}r Sonnenphysik}
\newacro{KPNO}{Kitt Peak National Observatory}
\newacro{LASP}{Laboratory for Atmospheric and Space Physics}
\newacro{LC}{liquid cristal}
\newacro{LMSAL}{Lockheed Martin Solar and Astrophysics Labratory}
\newacro{LOS}{line of sight}
\newacro{LTE}{local thermodynamic equilibrium}
\newacro{MC}{magnetic concentration}
\newacro{MCAO}{multi-conjugate adaptive optics} 
\newacro{MDI}{Michelson Doppler Imager}
\newacro{ME}{Milne-Eddington} 
\newacro{MHD}{magnetohydrodynamics}
\newacro{MOMFBD}{Multi-Object Multi-Frame Blind Deconvolution}
\newacro{MPE}{Max--Planck--Institut f\"ur extraterrestrische Physik}
\newacro{MPG}{Max--Planck--Gesellschaft}
\newacro{MPS}{Max Planck Institute for Solar System Research}
\newacro{MSSL}{Mullard Space Science Laboratory}
\newacro{MTF}{modulation transfer function}
\newacro{NAOJ}{National Astronomical Observatory of Japan}
\newacro{NASA}{National Aeronautics and Space Administration}
\newacro{NIST}{National Institute of Standards and Technology}
\newacro{NLTE}{non-local thermodynamic equilibrium}
\newacro{NLFFF}{non-linear force-free field}
\newacro{NOAA}{National Oceanic and Atmospheric Administration}
\newacro{non-E}{non-equilibrium}
\newacro{NSO}{National Solar Observatory}
\newacro{NWO}{Netherlands Organisation for Scientific Research}
\newacro{PHE}{propagating heating event}
\newacro{PRD}{partial redistribution}
\newacro{PROBA2}{PRoject for OnBoard Autonomy}
\newacro{PSBE}{post Saha-Boltzmann extinction}
\newacro{PSF}{point spread function}
\newacro{QS}{quiet Sun}
\newacro{QSEB}{quiet-Sun Ellerman-like brightening} 
\newacro{RAL}{Rutherford Appleton Laboratory}
\newacro{RBE}{rapid blue-shifted excursion}
\newacro{R-MHD}{radiation hydrodynamics}
\newacro{rms}{root mean square}
\newacro{RMS}{root mean square}
\newacro{ROB}{Royal Observatory of Belgium}
\newacro{ROI}{region of interest}
\newacro{RRE}{rapid red-shifted excursion}
\newacro{RTE}{radiative transfer equation}
\newacro{RTSA}{Radiative Transfer in Stellar Atmospheres}
\newacro{SCF}{slender \CaIIH\ fibril}
\newacro{SE}{statistical equilibrium}
\newacro{SB}{Saha Boltzmann}
\newacro{SDO}{Solar Dynamics Observatory}
\newacro{SJI}{slit-jaw image}
\newacro{SLI}{slit image}
\newacro{SNR}{signal-to-noise ratio}
\newacro{SO}{Solar Orbiter}
\newacro{SoHO}{Solar and Heliospheric Observatory}
\newacro{SP}{Spectropolarimeter}
\newacro{SST}{Swedish 1-m Solar Telescope}
\newacro{SUMER}{Solar Ultraviolet Measurements of Emitted Radiation}
\newacro{SUFI}{Sunrise Filter Imager}
\newacro{SVD}{singular value decomposition}
\newacro{SVST}{Swedish Vacuum Solar Telescope}
\newacro{STX}{Solar Telescope X}
\newacro{THEMIS}{T\'{e}lescope H\'{e}liographique pour l'Etude du 
   Magn\'{e}tisme et des Instabilit\'{e} Solaires}     
\newacro{TR}{transition region}
\newacro{TRACE}{Transition Region and Coronal Explorer}
\newacro{TSI}{total solar irradiance}
\newacro{UT}{Universal Time}
\newacro{UV}{ultraviolet}
\newacro{VAULT}{Very high angular resolution ultraviolet telescope}
\newacro{VIRGO}{Variability of solar IRradiance and Gravity Oscillations}
\newacro{VTT}{Vacuum Tower Telescope}    
\newacro{XRT}{X-Ray Telescope}

\long\def\startignore #1\stopignore{}   

\def\rmit#1{{\it #1}}              
\def\etal{\rmit{et al.}}           
           
\def\eg{\rmit{e.g.,}}              
\def\cf{cf.}                       

\def\specchar#1{\uppercase{#1}}    
\def\specand{ and }                
\def\specand{\,\&\,}               
\def\AlI{\mbox{Al\,\specchar{i}}}  

\def\CI{\mbox{C\,\specchar{i}}}

\def\CaII{\mbox{Ca\,\specchar{ii}}}

\def\FeI{\mbox{Fe\,\specchar{i}}} 
\def\FeII{\mbox{Fe\,\specchar{ii}}}

\def\HI{\mbox{H\,\specchar{i}}} 
 
\def\Hmin{\hbox{{\rm H}$^{^{_-}}$}}      
\def\HeI{\mbox{He\,\specchar{i}}}

\def\MgI{\mbox{Mg\,\specchar{i}}} 
\def\MgII{\mbox{Mg\,\specchar{ii}}}

\def\MnI{\mbox{Mn\,\specchar{i}}}

\def\NaI{\mbox{Na\,\specchar{i}}}

\def\OIII{\mbox{O\,\specchar{iii}}}

\def\SiI{\mbox{Si\,\specchar{i}}} 
 
\def\SiIII{\mbox{Si\,\specchar{iii}}}

\def\Halpha{\mbox{H\hspace{0.1ex}$\alpha$}} 

\def\Lyalpha{\mbox{Ly$\hspace{0.2ex}\alpha$}}

\def\HeIDthree{\mbox{He\,\specchar{i}\,\,D$_3$}}

\def\NaD{\mbox{Na\,\specchar{i}\,D}}    

\def\NaID{\mbox{Na\,\specchar{i}\,\,D}}

\def\Dtwo{\mbox{D$_2$}}

\def\MgIb{\mbox{Mg\,\specchar{i}\,b}}

\def\CaIIK{\mbox{Ca\,\specchar{ii}\,\,K}}       
\def\CaIIH{\mbox{Ca\,\specchar{ii}\,\,H}}
\def\CaIIHK{\mbox{Ca\,\specchar{ii}\,\,H{\specand}K}}
\def\HK{\mbox{H{\specand}K}}

\def\CaIR{\mbox{Ca\,\specchar{ii}\,8542\,\AA}} 

\def\hk{\mbox{h{\specand}k}}

\def\level #1 #2#3#4{$#1 \; ^{#2} \mbox{#3} ^{#4}$}   

\def\rma{{\rm a}}     
  
\def\rmd{{\rm d}}  
\def\rme{{\rm e}}

 \def\rmH{{\rm H}} 
 \def\rmI{{\rm I}}

\def\rms{{\rm s}}

\def\tis{\!=\!\!}                          
\def\tapprox{\!\approx\!\!}                
\def\dif{\: {\rm d}}                       
\def\ep{\:{\rm e}^}                        
\def\lambdop{\hbox{$\bf \Lambda$}}         

